\documentclass[twocolumn, trackchanges]{aastex701}
\usepackage{amsmath}
\usepackage{graphicx}

\begin{document}

\title{Stellar flare detection and characterization in ZTF High-cadence light curves}

\author[orcid=0009-0003-0631-5281,sname='Aratrika Ghosh']{Aratrika Ghosh}
\affiliation{Department of Astronomy and Astrophysics, Tata Institute of Fundamental Research, Homi Bhabha Road, Mumbai 400005, India}
\email[show]{aratrika.ghosh@tifr.res.in}  

\author[orcid=0000-0003-2896-1471,gname=Hanasoge, sname='Sur America']{Shravan Hanasoge} 
\affiliation{Department of Astronomy and Astrophysics, Tata Institute of Fundamental Research, Homi Bhabha Road, Mumbai 400005, India}
\email{hanasoge@tifr.res.in}

\author[orcid=0000-0002-3168-0139, gname=Matthew Graham]{Matthew J. Graham}
\affiliation{Division of Physics, Mathematics and Astronomy, California Institute of Technology, Pasadena, CA 91125, USA}
\email{mjg@caltech.edu}

\author[orcid=0000-0003-2242-0244, gname=Ashish Mahabal]{Ashish Mahabal}
\affiliation{Division of Physics, Mathematics and Astronomy, California Institute of Technology, Pasadena, CA 91125, USA}
\email{aam@astro.caltech.edu}

\begin{abstract}

Stellar flares are brief, intense increases in brightness resulting from magnetic reconnection events in stellar atmospheres. In this paper, we have developed a pipeline to detect and characterize flares in short-duration (approximately 6.2-hour) light curves from the Zwicky Transient Facility (ZTF) Extended Deep-Drilling survey. This method involves robust detrending with Tukey’s biweight estimator, followed by sigma-based thresholding and visual scrutiny to identify flare candidates. It is applied to 28 million light curves, and we found 331 flaring events from 310 stars across evolutionary stages and spectral types. We estimated flare energies by assuming a blackbody emission model for 234 flares with reliable distance, and flare morphologies were classified based on flare modeling using the template by \citet{mendoza_llamaradas_2022}. We obtained a power law correlation between energy and duration with an index of 0.19 $\pm$ 0.01. We also find similar power law correlation between the flare energy and $T_{\rm decay}$, $T_{\rm rise}$ and FWHM. This study demonstrates the power of high-cadence ground-based surveys for statistical and morphological flare studies across a broad stellar sample.

\end{abstract}


\section{Introduction} 

Stellar flares are brightening events associated with the sudden release of magnetic energy in the stellar atmosphere, analogous to solar flares. These are thought to originate from magnetic reconnection events in the stellar atmosphere, leading to the emission of electromagnetic radiation across the spectrum, from radio waves to gamma rays \citep{kowalski_stellar_2024}. Understanding the origin and mechanisms of these flares is crucial for comprehending magnetic activity and energy transport processes in stellar coronae as well as their impact on the habitability of planets \citep{segura_effect_2010, chadney_effect_2017, atri_modelling_2017}.

The stochastic nature of the flare necessitates statistical analyses to characterize its properties and the underlying physical mechanism \citep{hilton_m_2010, davenport_kepler_2014}. To this end, numerous studies have been conducted across various stellar types to quantify flare occurrence rates, energy distributions, and temporal characteristics. \citet{davenport_kepler_2016} performed a statistical analysis on more than 800,000 flares from 4041 stars from the Kepler \citep{borucki_kepler_2010} catalog. \citet{maehara_superflares_2012} and \citet{shibayama_superflares_2013} studied the statistical occurrence rates of superflares on G dwarfs, and \citet{hawley_kepler_2014} performed a detailed study of flares on M dwarfs, particularly on GJ 1243. \citet{van_doorsselaere_stellar_2017} performed a detailed flare study across spectral types and evolutionary stages, finding flaring events in 653 giants and on A-type stars, extending the studies performed by \citet{balona_kepler_2012, balona_flare_2015}. This hints at different possible flaring mechanisms due to the lack of a convective zone in A-type stars and a possible surface dynamo for giants. \citet{pietras_statistical_2022} also performed a detailed analysis of stellar flares from the first three years of TESS \citep{ricker_transiting_2014}, detecting more than 140,000 flares.

Beyond statistical occurrence rates, detailed studies of flare morphology provide deeper insights into the physical processes driving flare evolution. \citet{seli_stellar_2025} performed a detailed study of flare morphologies across main-sequence stars, while \citet{bruno_detailed_2024} modeled flares using the template derived by \citet{mendoza_llamaradas_2022} to study individual flare components and pre-flare dips. \citet{yudovich_analyzing_2025} focused on late-phase flare morphologies, identifying a correlation between the peak and bump amplitudes. \citet{howard_no_2022} showed that short-cadence photometry reveals fine substructures within flares.

Most detailed flare studies have been significantly advanced with the advent of space-based telescopes, such as Kepler and TESS; however, ground-based telescopes provide an important complement to space-based missions. \citet{jackman_ngts_2020, jackman_stellar_2021} utilized Next Generation Transit Survey \citep[NGTS;][]{wheatley_next_2018} data to study flares on stars near the Galactic disk as well as on pre-main-sequence stars, providing valuable insights into flare occurrences in diverse stellar populations. In addition, the wide field Evryscope survey \citep{evryscope_2014}, an array of small telescopes, has also been used to detect and characterize superflares in cool stars \citep {howard_evryflare_2019}. More recently, \citet{voloshina_snad_2024} employed high-cadence observations taken by the Zwicky Transient Facility  \citep[ZTF;][]{bellm_zwicky_2018} to study flares in M-dwarfs. In comparison to other ground based surveys, ZTF observes much fainter stellar populations, particularly in the galactic plane, therefore probing a complementary region of parameter space. Thus, ZTF deep drilling observations provide an excellent opportunity to understand stellar activity in this less explored regimes.

In this paper, we leverage expanded high-cadence ZTF survey data to detect and characterize stellar flares across a wide range of stellar types. We focused on the short baseline data and conducted a statistical analysis of the physical flare properties. Additionally, we performed morphological studies by fitting flare light curves with the empirical flare template proposed by \citet{mendoza_llamaradas_2022}. This paper provides a catalog of 310 flaring stars near the galactic plane, along with an estimation of various flare properties, including energy, duration, amplitude, FWHM, rising and decaying timescales. We describe the ZTF data in section \ref{sec:data} while section \ref{sec:methods} presents our flare detection and filtering algorithm followed by flare modeling. Section \ref{sec:analysis} describes the spectral characterization of flare stars and estimation of flare energies. In section \ref{sec:discussion}, we discuss our results and we summarizes our main findings and conclude in section \ref{sec:summary}.

\section{\label{sec:data}Data Overview}

The Zwicky Transient Facility is a ground-based northern sky optical survey that utilizes the 48-inch Schmidt telescope at Palomar Observatory, having a 47 $deg^2$ field of view and a median depth of g $\sim$20.8 and r $\sim$20.6 mag. ZTF's typical cadence is around 2-3 nights, but it also conducts dedicated 3-night short cadence observations in selected fields. The ZTF Galactic Science Working Group performed a high-cadence survey of three Galactic plane fields \citep{vanderbosch_upcoming_2024}, one per night, with Galactic latitude $|b|$ less than 20° from 1 July to 3 July 2024, for approximately 6.2 hours in the r band.

For this work, we have obtained data from the ZTF Data Release 23 (DR23)\footnote{\href{https://irsa.ipac.caltech.edu/frontpage/}{https://irsa.ipac.caltech.edu/frontpage/}} archive through the ZTF query interface, selecting only observations with catflags = 0 and filtering out measurements affected by lunar or cloud contamination. Since ZTF does not separate high-cadence data, a second layer of filtering involved retaining only sources observed during the high-cadence campaign. This step yielded approximately 28 million light curves, which were then processed through the flare-detection algorithm discussed below. To supplement the photometric time series, ZTF data were cross-matched with Gaia DR3 \citep{prusti_gaia_2016, vallenari_gaia_2023} to obtain stellar parameters (e.g., effective temperature, parallax, colors). The Bayestar19 3D dust map \citep{green_3d_2019, schlafly_measuring_2011} was also used to obtain distance-dependent reddening estimates across the sky, and Gaia EDR3 \citep{brown_gaia_2021, bailer-jones_estimating_2021} provided the geometric distances.

\section{\label{sec:methods}Methods}

\subsection{Flare Detection}

\begin{figure*}[t!]
    \plotone{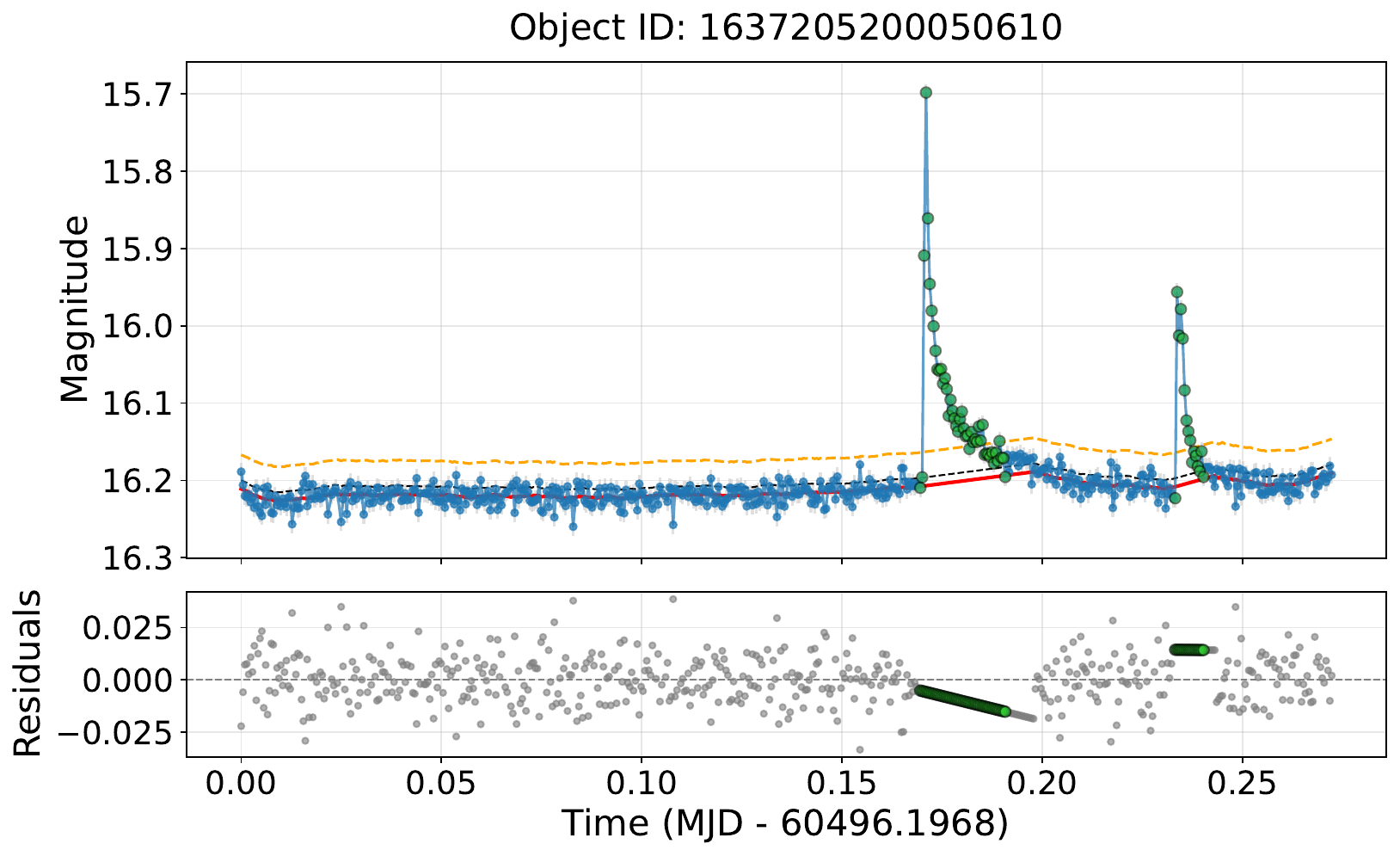}
    \caption{Flare detection method on ZTF Object ID: 1637205200050610 which is an M type star at $52.48\pm0.11$ pc. Top panel: Blue points represent the data, with the red line indicating the biweight estimated trend. Green points are possible flaring events. The black dashed line represents $1\sigma$ level, and the yellow dashed line represents $4\sigma$ level. Bottom panel: Grey points represent the residual after fitting the trend, and the green points represent the flaring event.}
    \label{fig:method}
\end{figure*}

\begin{figure*}[t]
    \gridline{
        \fig{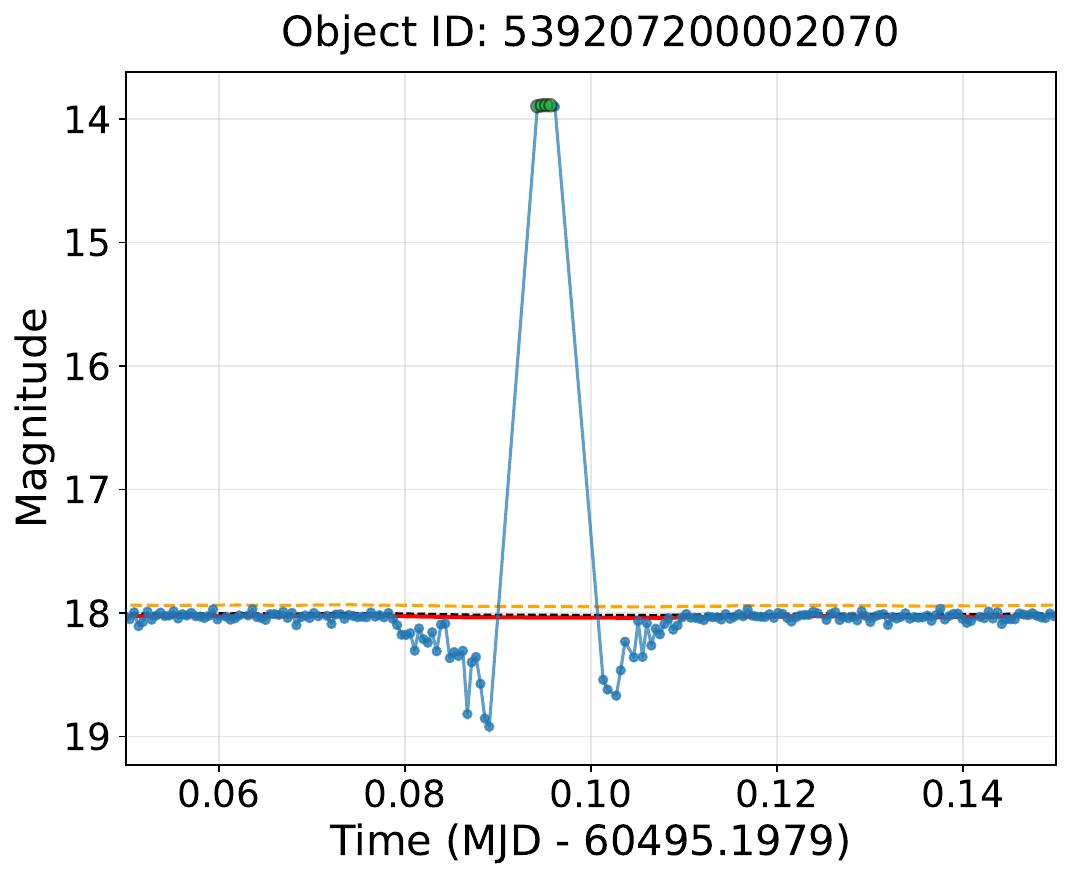}{0.43\textwidth}{}
        \fig{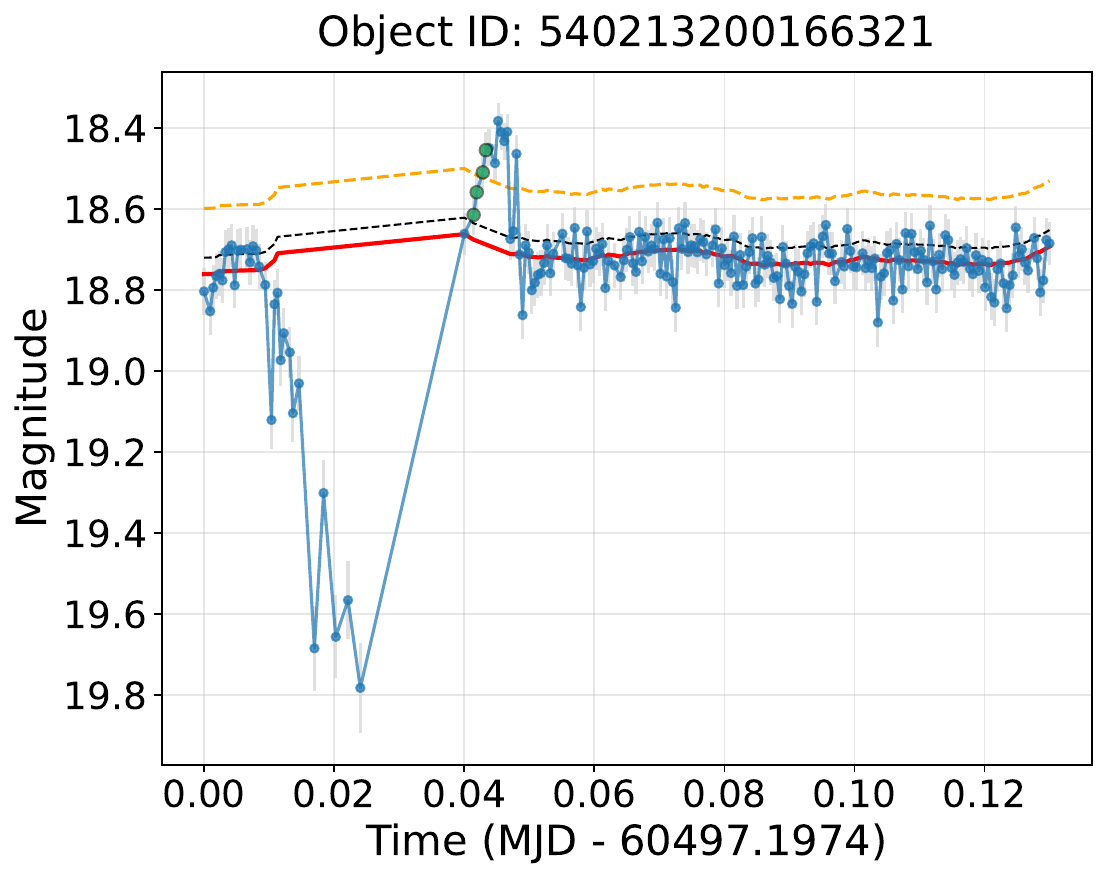}{0.43\textwidth}{}
    }
    \gridline{
        \fig{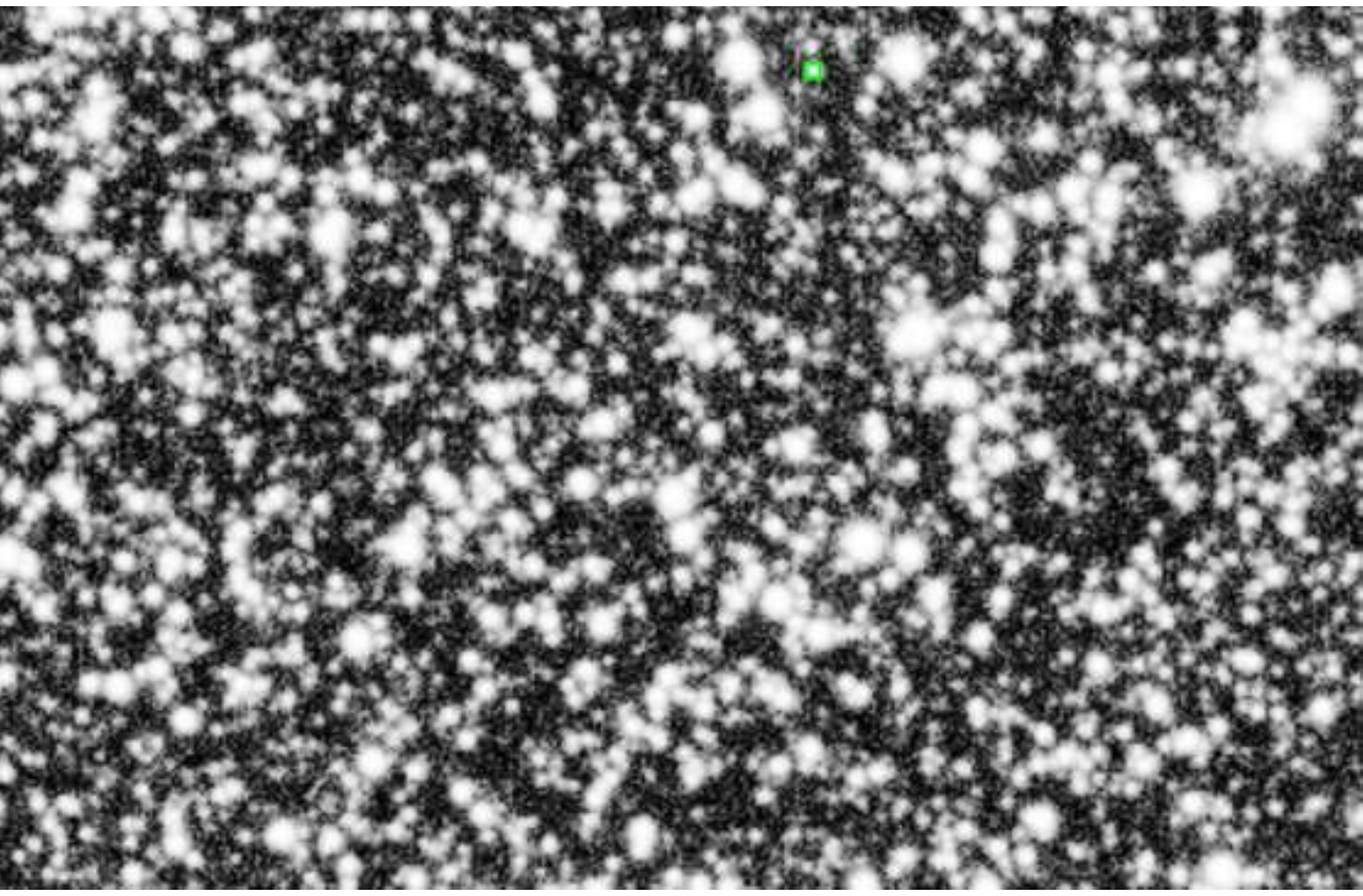}{0.28\textwidth}{(a)}
        \fig{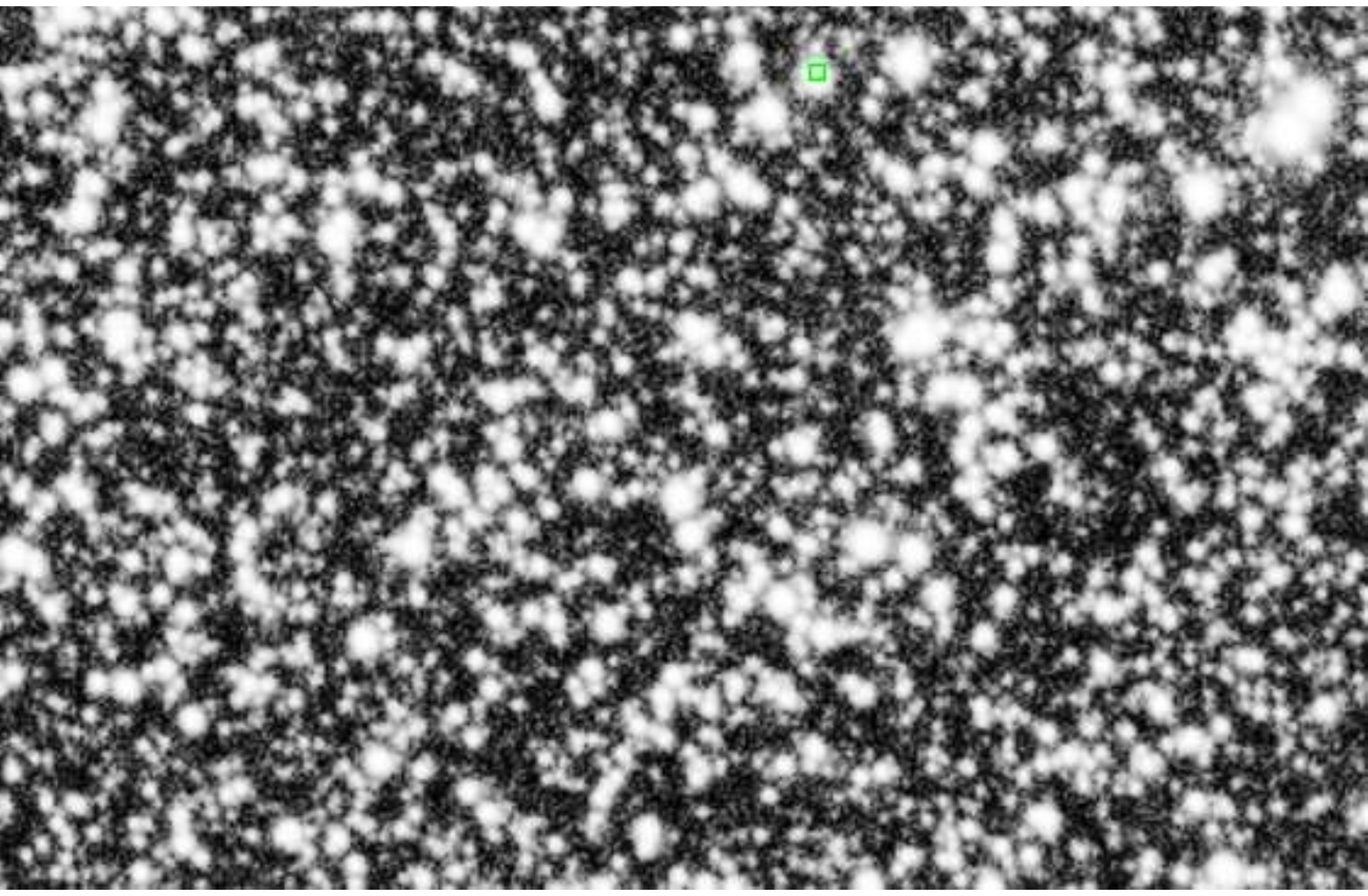}{0.28\textwidth}{(b)}
        \fig{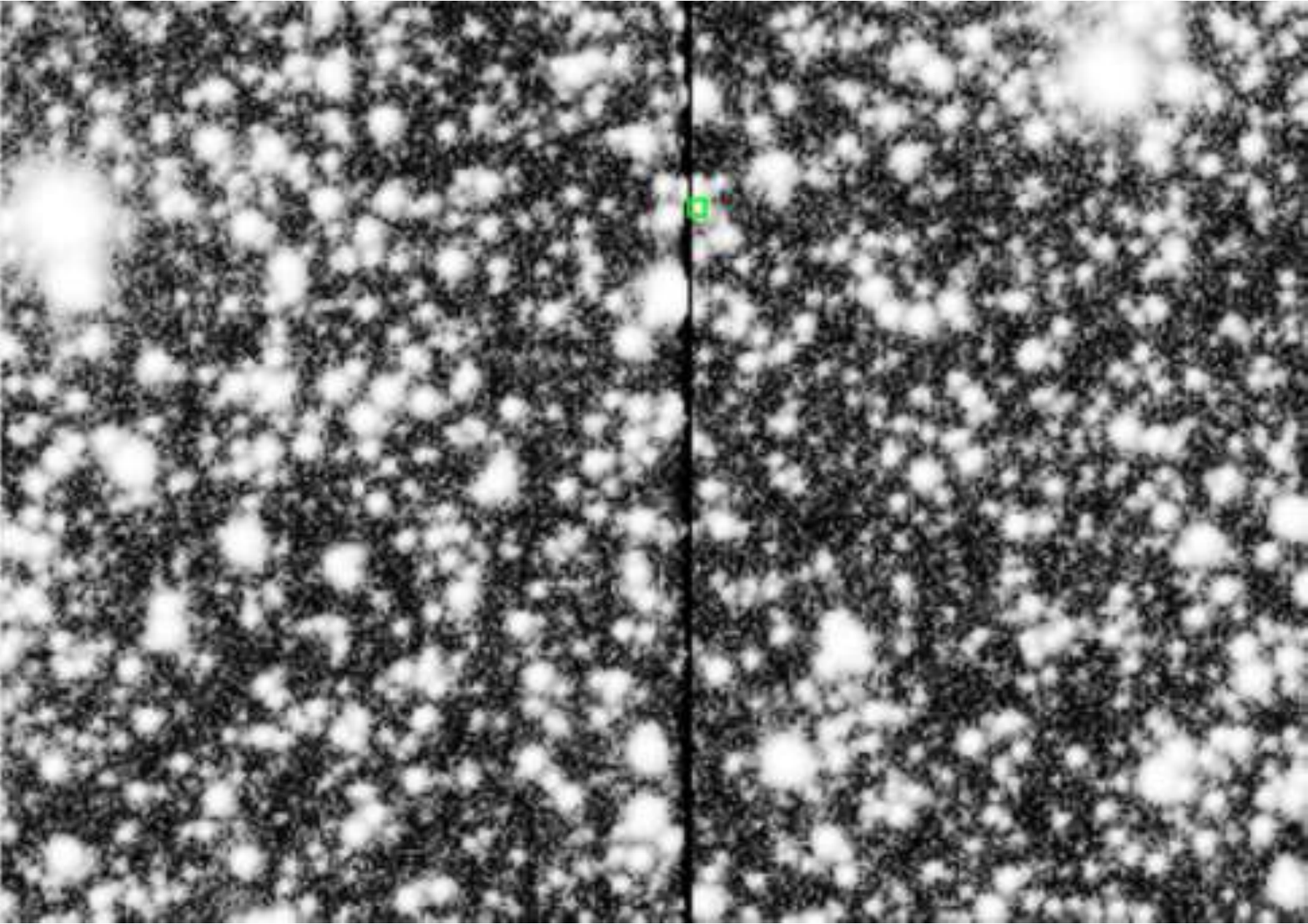}{0.27\textwidth}{(c)}
    }
    
    \caption{Examples of systematics which were identified by the algorithm as flares. Top: Asteroid trails giving rise to a plateau-like structure. The two FITS file (a) and (b) in the bottom show the asteroid passing the field of view giving rise to plateau like brightness. Bottom: The left panel shows the light curve with a dimming followed by brightening is marked as flare by the detection method. But the FITS file (c) show that it is due to CCD artifacts due to bad CCD line read. }
    \label{fig:systematics}
\end{figure*}

Stellar flares are sharp brightening events in photometric light curves and may be treated as statistical outliers in an otherwise smoothly varying light curve. A common approach is to remove all possible long-term variability and then identify significant outliers from the trend using a rolling window estimator. Typical choices for the trend estimator include a sliding mean, median filter, or Gaussian Processes \citep{vasilyev_superflares_2022, ilin_flares_2021}. However, these methods can be biased by strong outliers or might not scale well with such a large dataset, or produce a large number of false detections due to stellar variability. Another approach involves Bayesian analysis \citep{pitkin_bayesian_2014, gunther_stellar_2020} and various machine learning algorithms \citep{vida_finding_2018, lin_scalable_2024}, including convolutional neural networks \citep{feinstein_flare_2020} or recurrent neural networks \citep{vida_finding_2021}.

In this work, we have adopted rolling Tukey’s biweight location estimator \citep{mosteller_data_1977}, also known as an M-estimator, to perform robust detrending. This estimator downweights data points that are far from the central tendency, rather than assigning equal weight to all data points, as is the case with the rolling mean. This results in a smoother and more reliable baseline in the presence of high-amplitude transient events.

The biweight location estimator in window W is given as

\begin{equation}
\scalebox{0.9}{$
B =
\begin{cases}
m_{\mathrm{med}} +
\dfrac{
\sum\limits_{i \in W} u_i^2 (1 - u_i^2)^2 (m_i - m_{\mathrm{med}})
}{
\sum\limits_{i \in W} u_i^2 (1 - u_i^2)^2
}, & |u_i| < 1, \\[6pt]
0, & |u_i| \geq 1,
\end{cases}
$}
\label{eq:B_def}
\end{equation}
\begin{equation}
u_i = \frac{m_i - m_{\mathrm{med}}}{c\,\mathrm{MAD}}.
\label{eq:u_def}
\end{equation}

Here, $m_i$ are the magnitude values in the window, $m_{med}$ is the window median, MAD is the median absolute deviation, and c is a tuning constant (we have adopted c = 6).

\subsubsection{Detrending and Preprocessing}

The high-cadence observations have uneven sampling in the data, with cadences varying from 40 seconds to roughly 40 minutes in their 6-hour sequences. Hence, finding an optimal window width is crucial for accurately modeling the trend, and to account for this, we have used an adaptive rolling window estimator. 

We removed the upper and lower fourth percentiles of the magnitude values to eliminate extreme outliers. We then performed robust detrending using Tukey’s biweight location estimator from Astropy \citep{robitaille_astropy_2013, collaboration_astropy_2018, collaboration_astropy_2022}, initially applying a 30-minute sliding window (approximately 45 data points). If a given window contained fewer than five data points, the window size was iteratively expanded by a factor of 1.5 until at least five points were available or the window width reached a maximum of one hour. After this first-pass detrending, the standard deviation $\sigma$ was estimated by masking the fourth percentile tails of the detrended estimate. We then applied a 2.5$\sigma$ clipping on the detrended data to remove moderate outliers that could bias the trend estimation. On this sigma-clipped light curve, the detrending procedure was repeated twice to obtain the final trend, and the residual was provided with a new robust estimate of $\sigma$.

\subsubsection{Thresholding and False Detection Filtering}
\label{sec:detect}

To obtain flare candidates after trend estimation, we have used the following conditions: i) there must be at least three consecutive data points above the 4$\sigma$ level, and ii) there must not be any gap of more than 1 hour between these data points, and iii) the first data point above the threshold must not be the start of the light curve, and last data point above the threshold must not be the end of the light curve. The first criterion prevents any misdetection of cosmic ray hits as potential flares, the second criterion prevents misdetection of instrumental artifacts, and the third criterion rejects any incomplete flares. Once the potential flares are identified, the sequence is expanded on either side to obtain the duration of the flare, which is defined as the points for which the magnitude is outside the 1$\sigma$ level. 

Applying this method across the $\sim$28 million light curves yielded approximately 40,000 candidate flaring events, which were retained for further analysis. Due to the proximity of the observed fields to the Galactic plane, there is a significant blending in the ZTF images due to high stellar density. As a result, a significant fraction of the initially detected ``flares" are not actual astrophysical events but instead arise from contamination due to nearby stars. These were removed by ensuring that the number of outliers in a given light curve is no more than five, as blended light curves often appear spike-filled. This filtering reduced the number to approximately 2,700 candidates. 

Figure \ref{fig:method} shows the flare detection algorithm for a multi-flare event, and Figure \ref{fig:systematics} represents a few examples of false positives that were classified as flaring events by the detection algorithm. Visual inspection was carried out on the remaining candidates to remove multiple artifacts, including ghost reflections caused by bright stars, CCD lines with flux drops, transiting asteroids, and edge-of-the-camera effects. Moreover, due to pre-masking, false positives also include dips of variable stars. Symmetrical flare shape candidates potentially caused by stellar hotspots, self-lensing, or flares from other stars were also manually removed, resulting in the retention of 331 flaring events from 310 stars.

\subsection{Flare Modeling}
\label{sec:model_fl}

\begin{figure*}[t!]
    \plotone{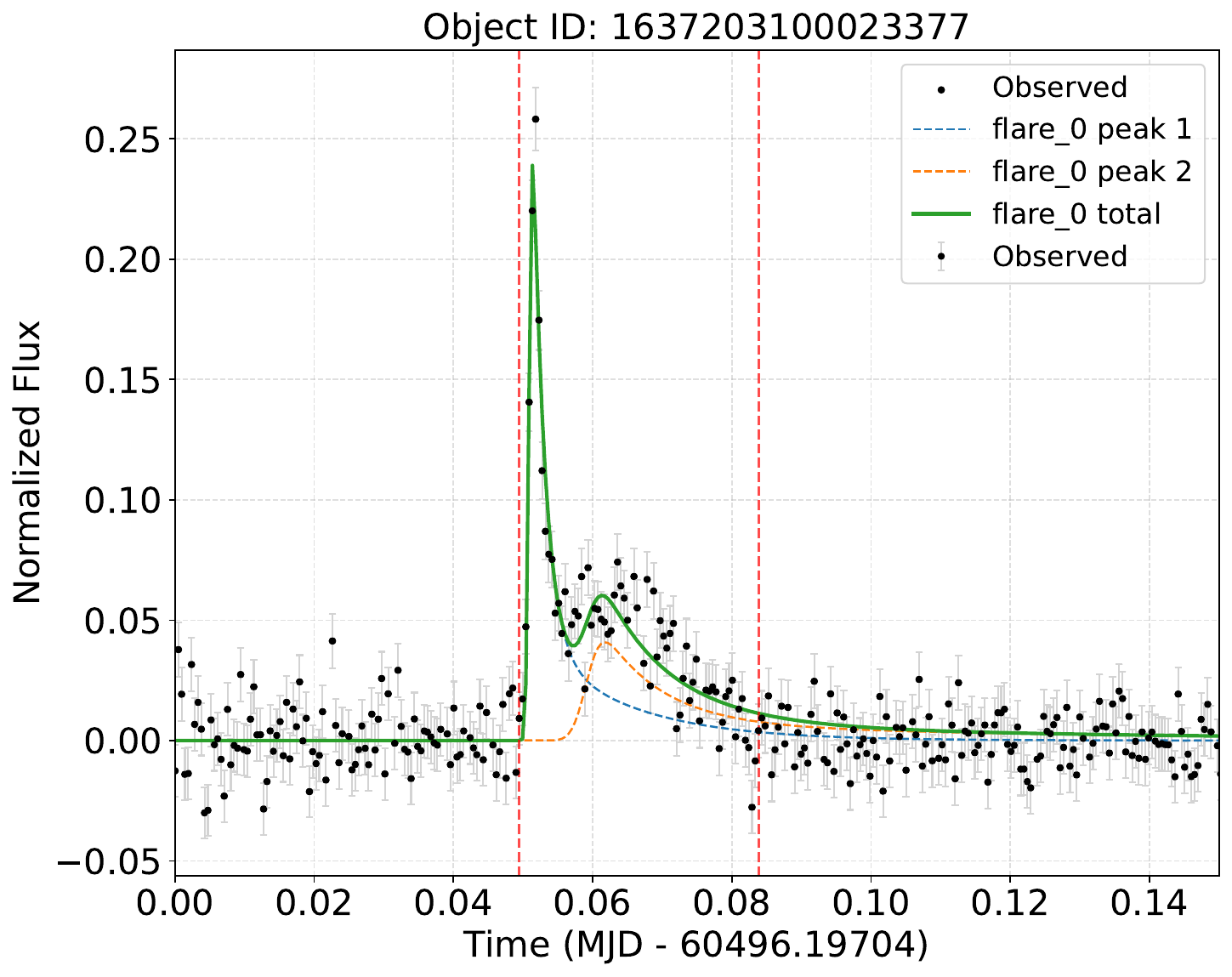}
    \caption{This figure shows a flare represented as $flare\_0$ modeled using the template from \citet{mendoza_llamaradas_2022}. Different colors indicate individual components of each flare. The total flare model (represented by the solid green line) is the sum of components of each flare. The red vertical dashed lines represent the start and the end of the flare. The start of the flare from the model is defined as the time when the flux first becomes non-zero whereas the end is defined as the point when flux becomes one-eighth of its peak value.}
    \label{fig:model}
\end{figure*}

Flares exhibit a wide range of structures, from simple single-peaked flares to complex flares with multiple peaks, including peak-bump, flat-top, and quasi-periodic pulsations (QPP) \citep{howard_no_2022, yang_properties_2023, inglis_quasi-periodic_2015, Ramsay_2021, yudovich_analyzing_2025, kowalski_stellar_2024, kowalski_time-resolved_2013}. These shapes encode the information for the underlying mechanisms driving magnetic reconnection and energy release. In a peak-bump flare, the first peak typically has a higher amplitude than the secondary peak, but there are cases where the peak in the gradual decay phase can also have a higher amplitude. The flat-top flare has a constant maximum amplitude at its peak, which could be due to a prolonged heating mechanism. In contrast, QPPs are characterized by periodic bumps in the decay phase, and the underlying mechanism remains unclear. The presence of such structures could be due to either the flare-cascade scenario, where multiple flares occur in the same active region, or to sympathetic flaring across multiple active regions \citep{davenport_kepler_2014, hawley_kepler_2014}. While the analytical models do not capture the entire structural complexity observed in high-cadence data, they still provide a way to test how well these templates can represent the observed diversity.

In this work, we have utilized the template developed by \citet{mendoza_llamaradas_2022}, which can describe a diverse range of structures. The template is constructed as a Gaussian heating pulse convolved with a sum of two exponential decay functions,

\begin{equation}
\scalebox{1.0}{$
f(t) = \frac{\sqrt{\pi} A C}{2}
\left[ F_{1} h(t,B,C,D_{1}) + F_{2} h(t,B,C,D_{2}) \right],
$}
\end{equation}
where
\begin{equation}
\scalebox{1.0}{$
h(t,B,C,D) =
\exp\!\left[-Dt + \left(\frac{B}{C} + \frac{DC}{2}\right)^{2}\right]
\, \mathrm{erfc}\!\left(\frac{B - t}{C} + \frac{DC}{2}\right)
$}
\end{equation}
where erfc is the complementary error function and f(t), the flux above the quiescent stellar baseline, depends on few parameters: i) $A$, the normalized peak of the flare, ii) $B$, the position of the peak, iii) $C$, the rising-time scale, iv) $D$, the cooling rate with $D_1$ and $D_2$ representing the fast- and slow-cooling timescales respectively, and v) $F_1$ and $F_2$ the relative cooling strengths. The time axis is shifted and rescaled to $(t-B)/$FWHM, where FWHM is the full width at half maximum of the flare peak. 

Model fitting was performed using nonlinear least-squares fitting with the Trust Region Reflective algorithm of \texttt{SciPY} \citep{virtanen_scipy_2020} on $A$, $B$, and FWHM. Rescaling the time axis standardized the heating and cooling timescales across flares, thus allowing the remaining parameters ($C$, $D_1$, $D_2$, $F_1$ and $F_2$) to be fixed to the values obtained from the MCMC fit in \citet{mendoza_llamaradas_2022}. The initial parameter guesses were obtained using the flare detection algorithm discussed in Section \ref{sec:detect}. The residual was obtained by subtracting the biweight estimator trend from the algorithm.

The validity of the fit is evaluated using reduced chi-square statistics. For each validated fit, we allowed multi peak modeling by fitting up to four flare peaks. We place an upper limit on the number of peaks to be considered in order to prevent overfitting to noise, which becomes increasingly important in short-cadence data. For statistical justification of adding multiple peaks, we required that the Akaike Information Criterion (AIC) for the fit with additional peaks be at least six units lower than that of the previous fit. This ensures that additional peaks are fitted only when there is statistical evidence that the simpler model is not well supported by the data. Figure \ref{fig:model} shows a complex flare fitted using the above procedure.

\section{\label{sec:analysis}Analysis}

\subsection{Star Characterization}

As spectral information is unavailable for most of the stars in our sample, we have used photometric color and absolute magnitude to classify stars across evolutionary stages and spectral types. We corrected the colors for galactic reddening. We distinguish between the main-sequence stars, subgiants, giants, and white dwarfs using the method described by \citet{daltio_dissecting_2021} in their Figure 7.

Main sequence stars are categorized into three types, namely Hot Main Sequence stars with $M_G \leq 4 \text{ and } -0.25 \leq BP - RP \leq 1$, Cold Main sequence stars I with $4 \leq M_G \leq 10 \text{ and } BP - RP \geq 0.5$, and Cold Main sequence stars II with $ M_G \leq 4 \text{ and } BP - RP \geq 1.5$, which targets the reddest dwarfs. For Subgiants and Red Giants,  $M_G < 4 \text{ and } BP - RP > 1$ and finally for white dwarfs, we have $M_G > 10 \text{ and } BP -RP < 0.8$. 

Furthermore, main-sequence stars were sub-classified into spectral types M, K, G, F, and A based on their BP-RP colors, following Table 5 of \citet{pecaut_intrinsic_2013}.

Applying this classification scheme to our sample, we obtained the following distributions for stellar luminosity class:

\begin{itemize}
    \item \textbf{Main Sequence:} 284 stars
    \item \textbf{Giants:} 7 stars
    \item \textbf{Unknown class} (due to missing \( M_G \) and \( BP - RP \)): 18 stars
\end{itemize}

For spectral types among the main-sequence stars, the distribution is:

\begin{itemize}
    \item \textbf{M type:} 196 stars
    \item \textbf{K type:} 62 stars
    \item \textbf{G type:} 15 stars
    \item \textbf{F type:} 6 stars
    \item \textbf{A type:} 3 stars
\end{itemize}

This classification shows that the flaring star samples are dominated by M dwarfs, which is consistent with the expected higher flare rate in low-mass stars as in \citet{gunther_stellar_2020}.

\subsection{Flare Energy Estimation}
\label{sec:energy}

To calculate the flare energy, we have assumed an optically thick blackbody with a temperature of $T_{fl}=10,000$ K in the active region of the stellar surface. This simplification is widely applied in the literature in the analysis of single-band observations \citep{hawley_x-ray--heated_1992}. In reality, the temperature changes throughout the flare evolution, and a blackbody model also cannot correctly explain the Balmer jumps in the near-UV region. Thus, the flare energy obtained should be regarded as a lower limit. We have calculated the energy according to the prescription outlined by \citet{voloshina_snad_2024}.
The bolometric luminosity of the flare is:
\begin{equation}
    L_{fl}(t) = \sigma_BT^4_{fl}\mathcal{A}_{fl}(t) \text{,}
\end{equation}
where $\mathcal{A}_{fl}(t)$ is the area of the active region that changes over time, which is estimated using the projected area on the image plane, $A_{\perp}$.

Given the spectral flux density of the flare as $F_{\nu} = \Omega B_{\nu}$, where $\Omega = \frac{A_{\perp}}{d^2}$ is the angle subtended by the flare at distance $d$ obtained using the Gaia EDR3 catalog and $B_{\nu}$ is the blackbody intensity. Now, we obtain flux density averaged over the transmission function of ZTF's r-band filter\footnote{ZTF $r$-band filter from the SVO Filter Profile Service: \href{https://svo2.cab.inta-csic.es/theory/fps/index.php?id=Palomar/ZTF.r\&\&mode=browse\&gname=Palomar\&gname2=ZTF\#filter}{SVO ZTF r-band filter}} \citep{koornneef_synthetic_1986}. Hence,
\begin{align}
F_r &= 
\frac{\int \lambda F_{\lambda}(t)R(\lambda)d\lambda}{\int \lambda R(\lambda)d\lambda} \\
&= 
\frac{ \mathcal{A}_\perp }{ d^2 } 
\cdot
\frac{\int \lambda B_{\lambda}(T_{fl})R(\lambda)d\lambda}{\int \lambda R(\lambda)d\lambda},\\
&=\frac{ \mathcal{A}_\perp }{ d^2 }B_r(T_{fl}),
\end{align}
where $B_r(T_{\mathrm{fl}})$ (erg cm$^{-2}$ s$^{-1}$ \AA$^{-1}$) is the average blackbody intensity in the ZTF $r$-band, and $R(\lambda)$ is the $r$-band transmission function.

Therefore, the projection area is
\begin{equation}
    A_{\perp} = d^2\frac{F_r(t)}{B_r(T_{fl})}.
\end{equation}

Finally, the bolometric luminosity is given as
\begin{equation}
    L_{fl}(t) = \sigma_BT^4_{fl} d^2 \frac{\mathcal{A}_{fl}(t)}{A_{\perp}}\frac{F_r(t)}{B_r(T_{fl})}.
\end{equation}

The area factor remains constant over time and is assumed to be unity. The final expression for the total energy of the flare: 
\begin{equation}
    E_{fl} = \int_{t \in t_{fl}} L_{fl}(t)dt
\end{equation}

Correcting for extinction $A_r$ using a 3D dust map and the reference AB-magnitude zero point, $m_0$, the flux is given as:
\begin{equation}
    f(t) = 10^{-0.4\left(m(t)-m_0 - A_r\right)}
\end{equation}
Here, $F_r(t) = f(t)-f^*$ where $f^*$ is the quiescent flux obtained from the median trend flux within the flare window.

\section{\label{sec:discussion}Discussion}

\begin{deluxetable*}{lcccccccc}
\tablecaption{Flare Properties from ZTF Light curves}
\label{tab:flare_table}
\tablehead{
    \colhead{Gaia ID} & \colhead{RA} & \colhead{Dec} & \colhead{Distance (pc)} & \colhead{Flare Amplitude $\frac{\Delta F}{F_{max}}$} & \colhead{Duration (min)} & \colhead{Energy (erg)}  &\colhead{ZTF OID\tablenotemark{a}}
}
\startdata
1834688502768999680 & 300.8964 & 25.2346 & $402.33^{+10.37}_{-12.69}$ & 0.19 & 115.30 & $7.18\substack{+0.38 \\ -0.47}\times10^{33}$ & 1637215300083415 \\
4313128174433089920 & 287.3892 & 12.1095 & $1531.02^{+317.38}_{-613.61}$ & 0.41 & 71.84 & $2.71\substack{+1.01 \\ -2.61}\times10^{35}$ & 539209300083780 \\
1831951229901066240 & 305.0655 &  24.1946 & $411.91^{+45.39}_{-43.86}$ & 0.37 & 21.67 & $7.35\substack{+1.57 \\ -1.68}\times10^{32}$ & 1637209200006835 \\
1823165792701775616 & 299.3774 &  19.5844 & $5838.17^{+1951.18}_{-2217.86}$ & 0.62 & 74.62 & $2.26\substack{+1.26 \\ -2.04}\times10^{36}$ & 1637204400084529  \\
4505456015324875520 & 281.9567 & 12.9843 & 5538.52$^{+1765.53}_{-3212.36}$ & 0.75 & 59.04 & $2.81\substack{+1.51 \\ -4.21}\times10^{36}$ & 539212200130867 \\
\enddata
\tablenotetext{a}{Lightcurves and FITS file can be accessed from \href{https://ztf.snad.space/}{https://ztf.snad.space/} using the ZTF OIDs.}
\tablecomments{Examples of a few ZTF flares from our sample. The full table, including all stellar and flare parameters, is available in machine-readable format. The full table includes the Gaia source identifier, sky coordinates, geometric distance and its uncertainties, effective temperature and its uncertainties, [Fe/H] and its uncertainties, flare amplitude, duration, FWHM, $T_{\rm rise}$, $T_{\rm decay}$, flare energy and its uncertainties, morphology classifications from both restricted and flexible model (Section \ref{sec:model_fl}, Appendix \ref{sec:app}), and ZTF object identifier. The \(T_{\rm eff}\) and \([\mathrm{Fe/H}]\) values are adopted from the Gaia DR3 GSP-Phot astrophysical parameter catalog \citep{Creevey_2023, andrae_gaia_2023, vallenari_gaia_2023}, while the distances are derived from Gaia EDR3 parallaxes. The quoted uncertainties on \(T_{\rm eff}\) and \([\mathrm{Fe/H}]\) correspond to the 16th and 84th percentiles of the GSP-Phot posterior distributions, while the lower and upper distance values correspond to the respective distance quantiles from \citet{bailer-jones_estimating_2021}.
\\
The machine readable table is available with the online article.
}
\end{deluxetable*}

From the initial sample of 310 light curves, we cross-matched with Gaia to obtain the photometric colors, distances, and other spectral information for estimating flare properties. We found that distance was available for 297 stars from the Gaia EDR3 catalog. To ensure reliable energy estimation, we imposed a limit of 33 percent uncertainty on distance measurement and finally obtained 234 events. This cutoff was chosen because it corresponds to an uncertainty of approximately 66 percent in energy. The uncertainty is less than one order of magnitude and is negligible compared to the orders of magnitude spanned by the flare energies, so it only has limited impact on the analysis. The stellar and flare properties of the ZTF sources are listed in Table~\ref{tab:flare_table}.

\subsection{Flare Amplitude, Duration and FWHM}

\begin{figure*}[ht!]
    \plotone{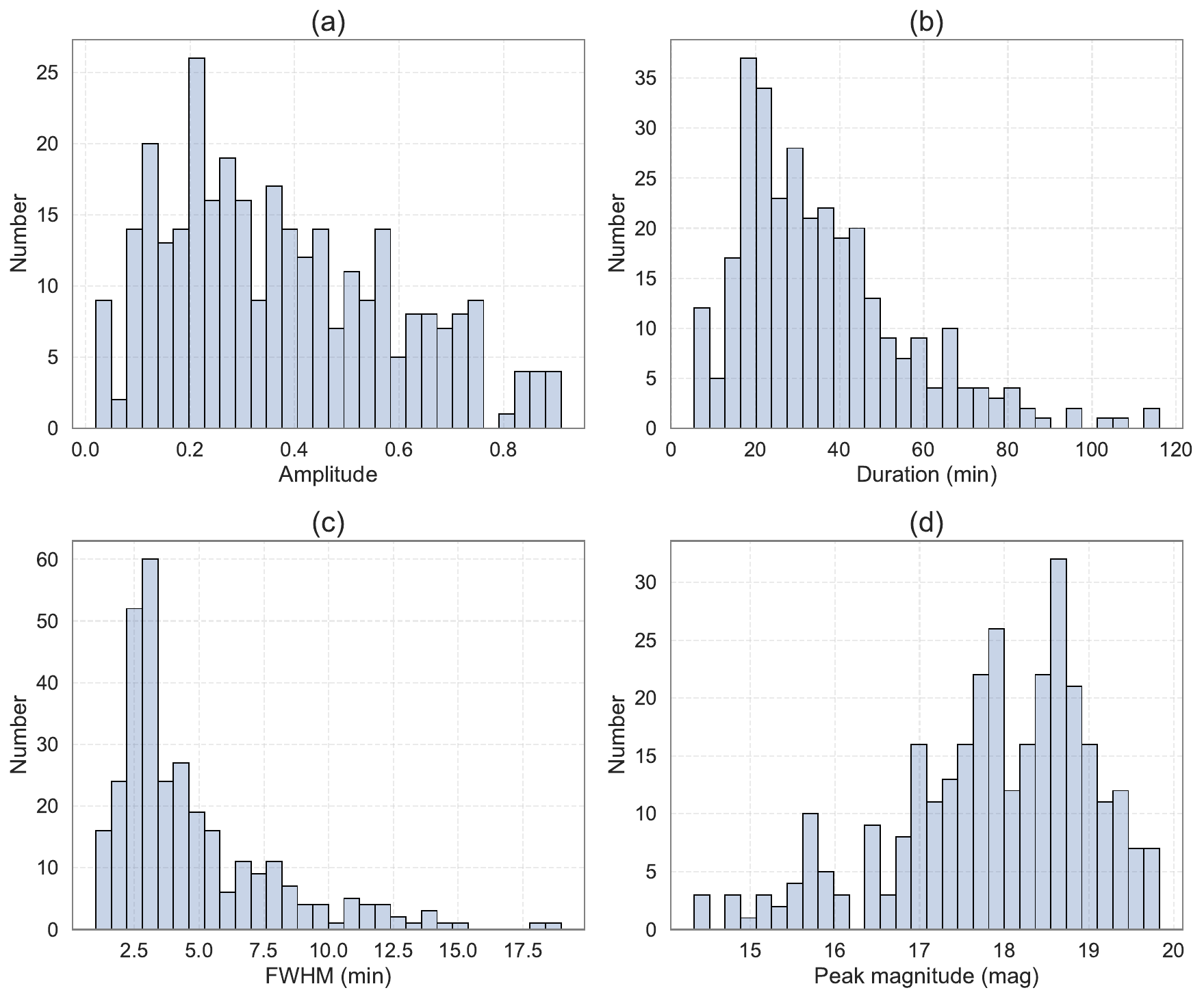}
    \caption{Top Left: Frequency distribution of flare amplitude. 
    Top Right: Frequency distribution of flare duration.
    Bottom Left: Frequency distribution of flare FWHM.
    Bottom Right: Frequency distribution of flare peak magnitudes.}
    \label{fig:hist1}
\end{figure*}

\begin{figure*}[t!]
    \plotone{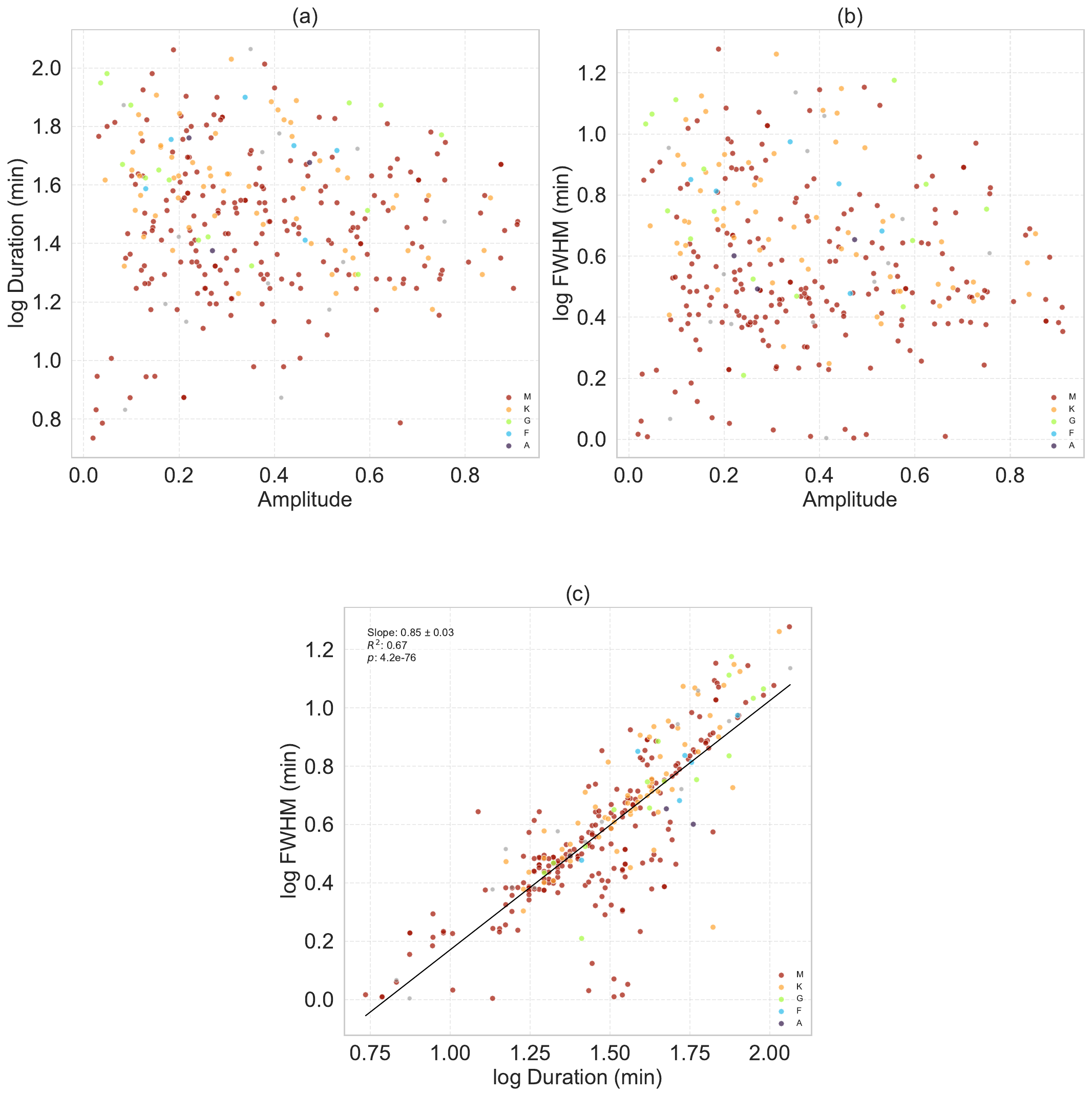}
    \caption{Relationship between flare duration, amplitude, and FWHM. Blue dots represent the M dwarfs, yellow dots represent the K dwarfs, green dots represent the G dwarfs, red dots represent the F dwarfs, purple dots represent the A-type stars, and black dots represent the remaining sample, mainly subgiants/red giants. Top Left: Relationship between log duration and flare amplitude. 
    Top Right: Relationship between flare amplitude and flare FWHM.
    Bottom: Relationship between log duration and log flare FWHM showing a strong positive correlation between the two quantities.}
    \label{fig:scatter1}
\end{figure*}

The flare amplitude is computed as $\frac{\Delta F}{F_{\rm max}}$ where $\Delta F$ is the change in the flare flux relative to the baseline. The distribution of flare normalized amplitude is given in Figure \ref{fig:hist1}(a), while the distribution of flare peak magnitude is shown in Figure \ref{fig:hist1}(d). The minimum and maximum relative amplitudes are 0.019 and 0.91, respectively. The flare duration is calculated as time duration of the flaring candidates identified using the method discussed in Section \ref{sec:detect}. Figure \ref{fig:hist1}(b) shows the distribution of flare duration with maximum and minimum durations of 5.43 minutes and 1.93 hours. The observed duration range is limited by the short baseline, causing long-duration flares to be treated as baseline and subsequently rejected by the detection algorithm. Figure \ref{fig:hist1}(c) shows the distribution of the flare FWHM calculated as the time interval over which the flux remains above the half maximum level relative to the baseline. We find that from Figure \ref{fig:scatter1}(a) there is no statistically significant correlation between flare amplitude and duration. Similarly, for flare amplitude and FWHM, there is no significant correlation between the two properties from Figure \ref{fig:scatter1}(b). However, there is a strong positive correlation between flare duration and FWHM from Figure \ref{fig:scatter1}(c), consistent with the intuition that broader flare profiles have longer duration.  

\subsection{Rise and Decay Timescales}

\begin{deluxetable*}{l c c c c c c}
\tablecaption{Flare Statistics for FWHM, $T_{\rm rise}$ and $T_{\rm decay}$ \label{tab:flare_stats1}}
\tablehead{
\colhead{Property} & \colhead{Minimum} & \colhead{Q1} & \colhead{Q3} & \colhead{Maximum} & \colhead{Mean}  & \colhead{Median}
}
\startdata
$T_{\rm rise}$ (min) & 0.66 & 2.03 & 4.09 & 16.27 & 3.51 & 2.72 \\
$T_{\rm decay}$ (min) & 4.77 & 18.99 & 41.18 & 103.10 & 32.56 & 28.50  \\
FWHM (min) & 1.01 & 2.60 & 5.66 & 18.97 & 4.67 & 3.51 \\
\enddata
\tablecomments{These quantities are calculated for all flare morphologies. $Q_1$ and $Q_3$ denote the first and third quartiles, respectively.}
\end{deluxetable*}

\begin{figure*}[t!]
    \plotone{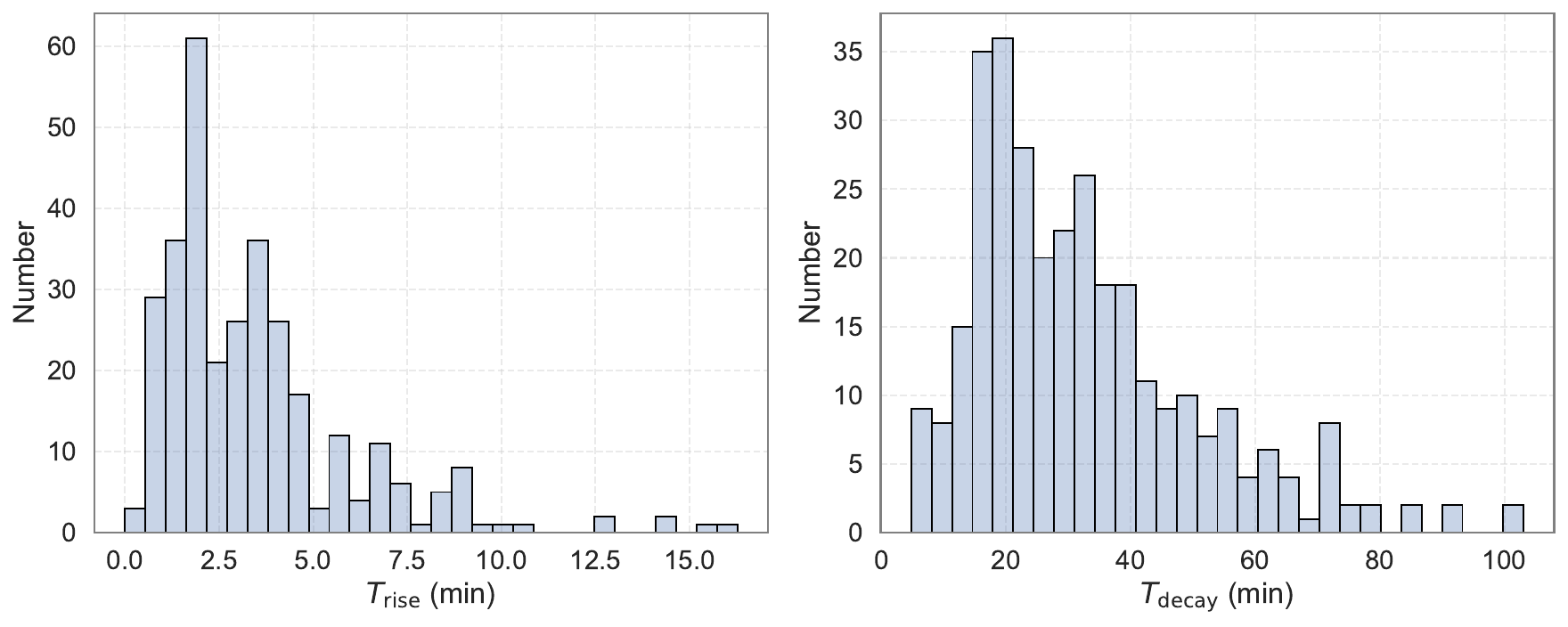}
    \caption{Left: Frequency distribution of $T_{\rm rise}$.
    Right: Frequency distribution of $T_{\rm decay}$.}
    \label{fig:hist2}
\end{figure*}

\begin{figure*}
        \plotone{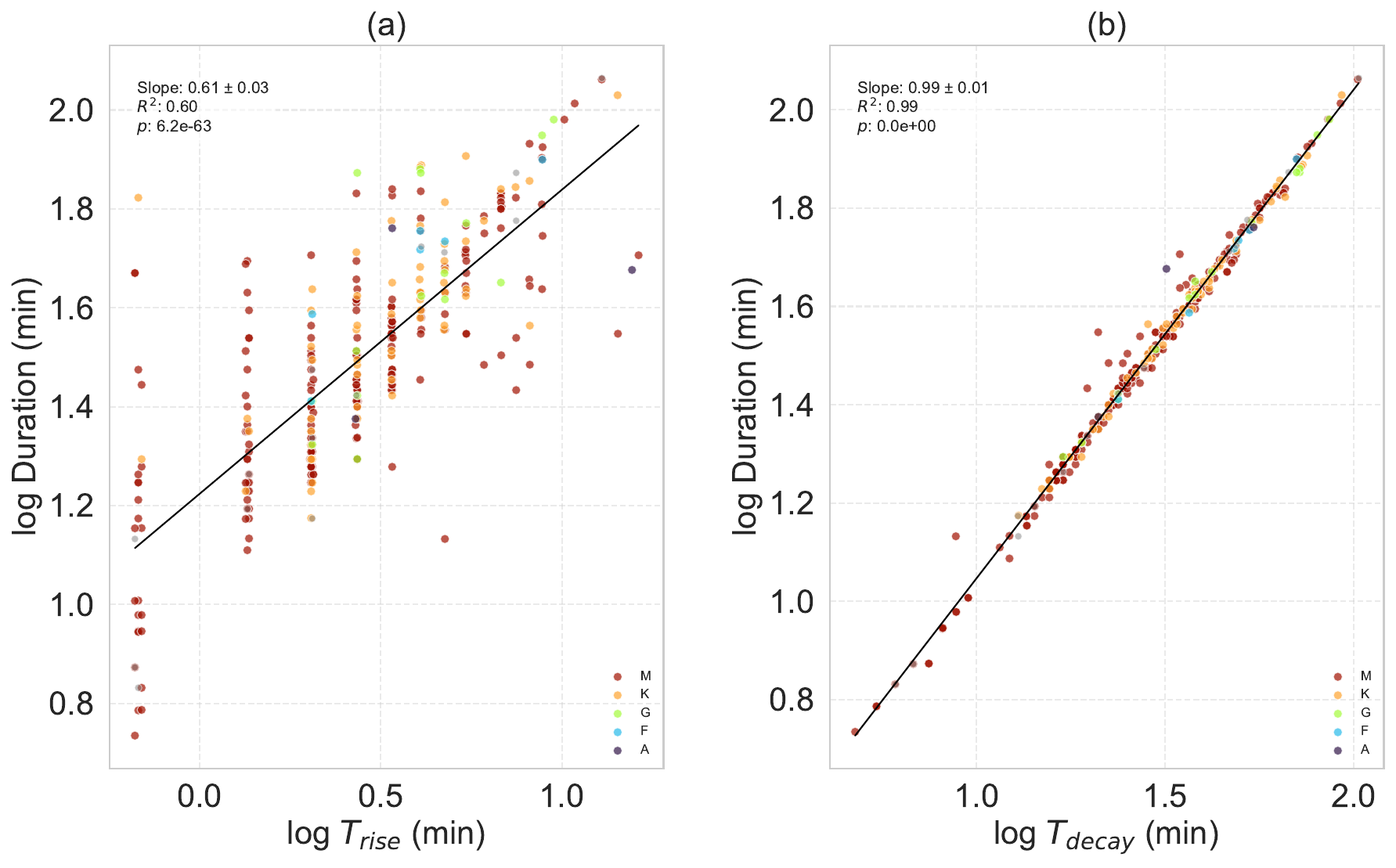}
    \caption{Relationship of flare duration with $T_{\rm rise}$ and $T_{\rm decay}$. Blue dots represent the M dwarfs, yellow dots represent the K dwarfs, green dots represent the G dwarfs, red dots represent the F dwarfs, purple dots represent the A-type stars, and black dots represent the remaining sample, mainly subgiants/red giants. Left: Relationship between $T_{rise}$ and log duration. Right: Relationship between $T_{\rm decay}$ and log duration, with the black line representing the best fit.}
    \label{fig:scatter2}
\end{figure*}

Table \ref{tab:flare_stats1} shows the statistical quantities for $T_{\rm rise}$ and $T_{\rm decay}$ for all the flaring events, which denote the time intervals from the start of each flare to its peak and from the peak to its end, respectively. Here, the peak of each flare is determined by identifying the data point with the maximum flux. We find that all the quantities of $T_{\rm decay}$ are significantly longer than those of $T_{\rm rise}$, which is consistent with previous studies \citep{yang_properties_2023}. Figure \ref{fig:hist2} shows the distributions for $T_{\rm rise}$ and $T_{\rm decay}$. This implies that the duration of the flare is dominated by the decay timescale as seen from the strong positive correlation between them in Figure \ref{fig:scatter2}. There is also a positive correlation between $T_{\rm rise}$ and log duration, but it is much weaker than that of $T_{\rm decay}$.

\subsection{Flare Energy}

\begin{figure}[]
        \plotone{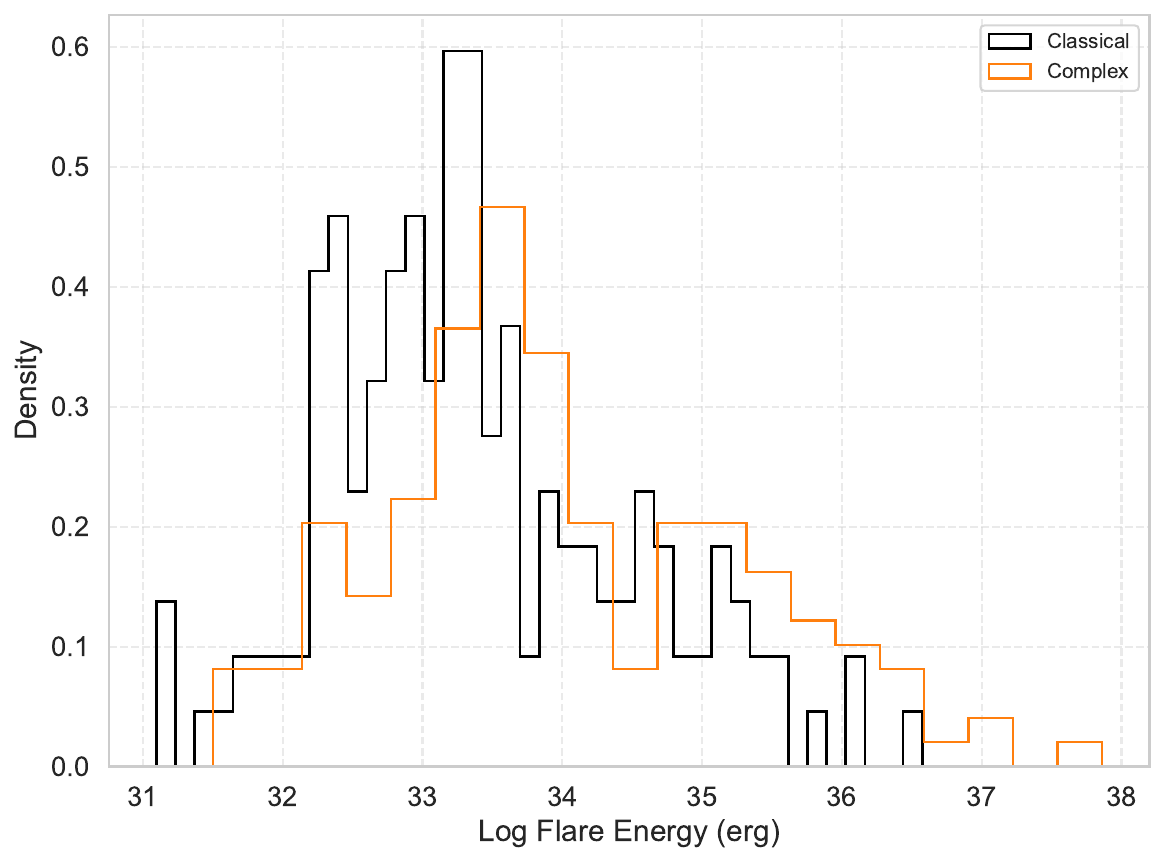}
    \caption{Distribution of log Energy for classical and complex(peak-bump and multi peak) flares where the morphology classification is performed following the criteria given in Section \ref{sec:morph}.}
    \label{fig:morph_en}
\end{figure}

Figure \ref{fig:morph_en} shows the distribution of the log energy of the flaring events for different morphological classes using the procedure describe in Section \ref{sec:morph}. Due to significant uncertainties in estimating distances, energy was calculated for only a subset of the stellar sample. The flare energies span a range from $1.34 \times 10^{31}$ erg to $7.25 \times 10^{37}$ erg with most concentrated around $10^{33}$ erg.

\begin{figure*}
    \plotone{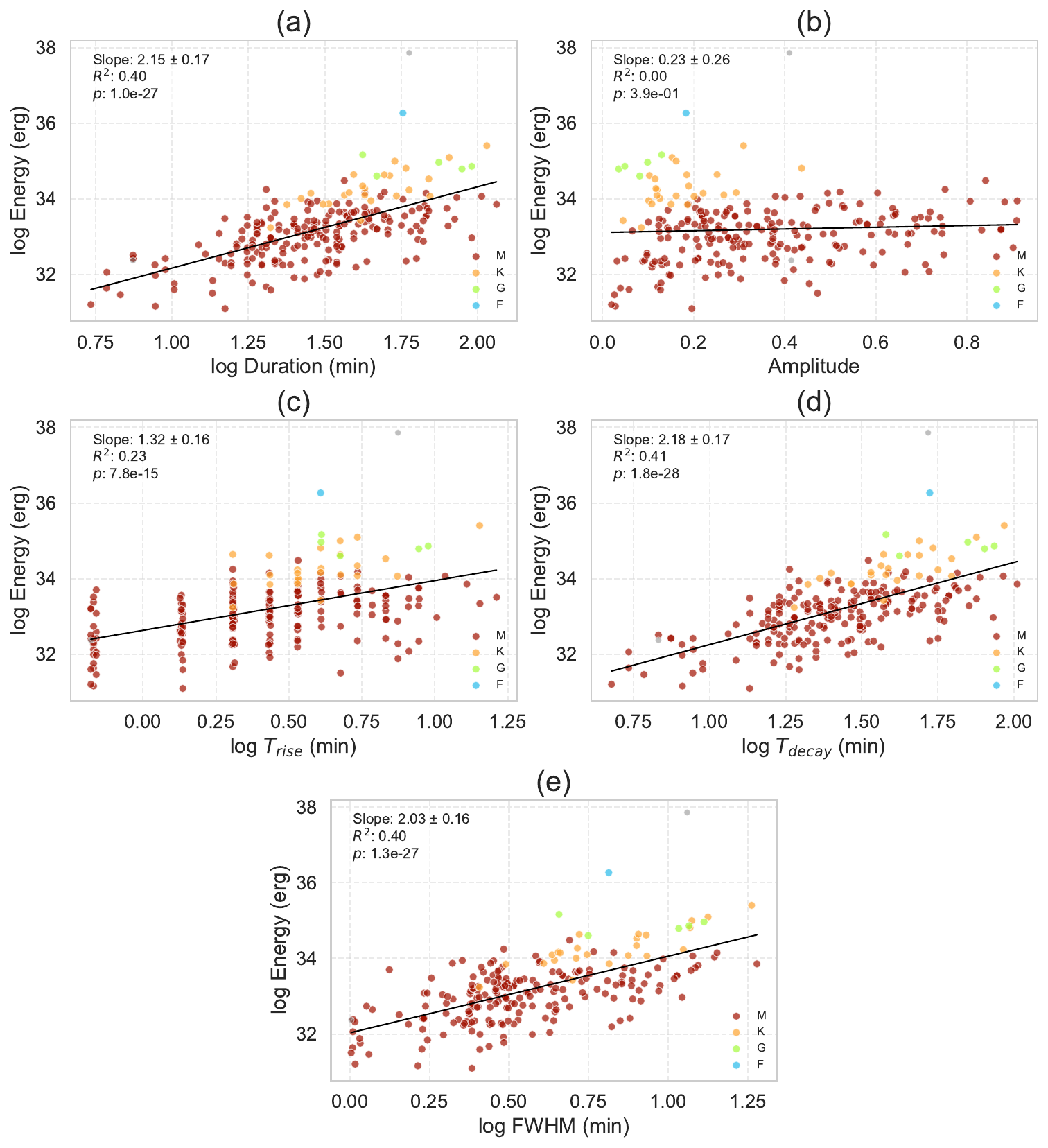}
    \caption{Relationships of flare energy with flare properties. Blue dots represent M dwarfs, yellow K dwarfs, green G dwarfs, red F dwarfs, and black dots represent subgiants/red giants. Panel (a) shows the connection between log energy and log duration, with the black line representing the best fit. Panel (b) shows the relationship between log energy and amplitude. Panel (c) tracks log energy vs the log decay timescale along with a best-fit line (slope 2.18). Panel (d) shows a weak correlation between log energy and rise timescale, and Panel (e) captures the behaviour of log flare energy vs FWHM}
    \label{fig:scatter3}
\end{figure*}

Figure \ref{fig:scatter3}(a) shows a strong positive linear correlation between log duration and log energy. Based on magnetic reconnection theory, \citep{maehara_statistical_2015} showed that a power law adequately related the two properties, given as $T_{\rm dur} \sim E^{\gamma}$ with $\gamma = \frac{1}{3}$. We have obtained $\gamma = 0.19 \pm 0.01$, which is lower than the values reported by the previous studies  \citep{yang_properties_2023, maehara_statistical_2015}. A similar positive correlation was found between log energy and $T_{\rm decay}$, which is strongly correlated with flare duration. No correlation was observed between log energy and amplitude as seen in Figure \ref{fig:scatter3}(b). For our sample, we find that the total energy of the flare events depends primarily on flare duration, which is consistent with the fact that highly energetic flares last longer. Figure \ref{fig:scatter3}(d) shows a similar power-law correlation between flare energy and decay timescale, implying that the cooling timescale dominates the flare duration. There is a positive correlation between flare energy and the heating timescale (Figure \ref{fig:scatter3}(c)), but it is much weaker than the cooling timescale. Figure \ref{fig:scatter3}(e) shows a weak positive correlation between log energy and FWHM, indicating that extended flares generally contribute more to the total energy output.

\subsection{Color-Magnitude Diagram}

\begin{figure}
        \plotone{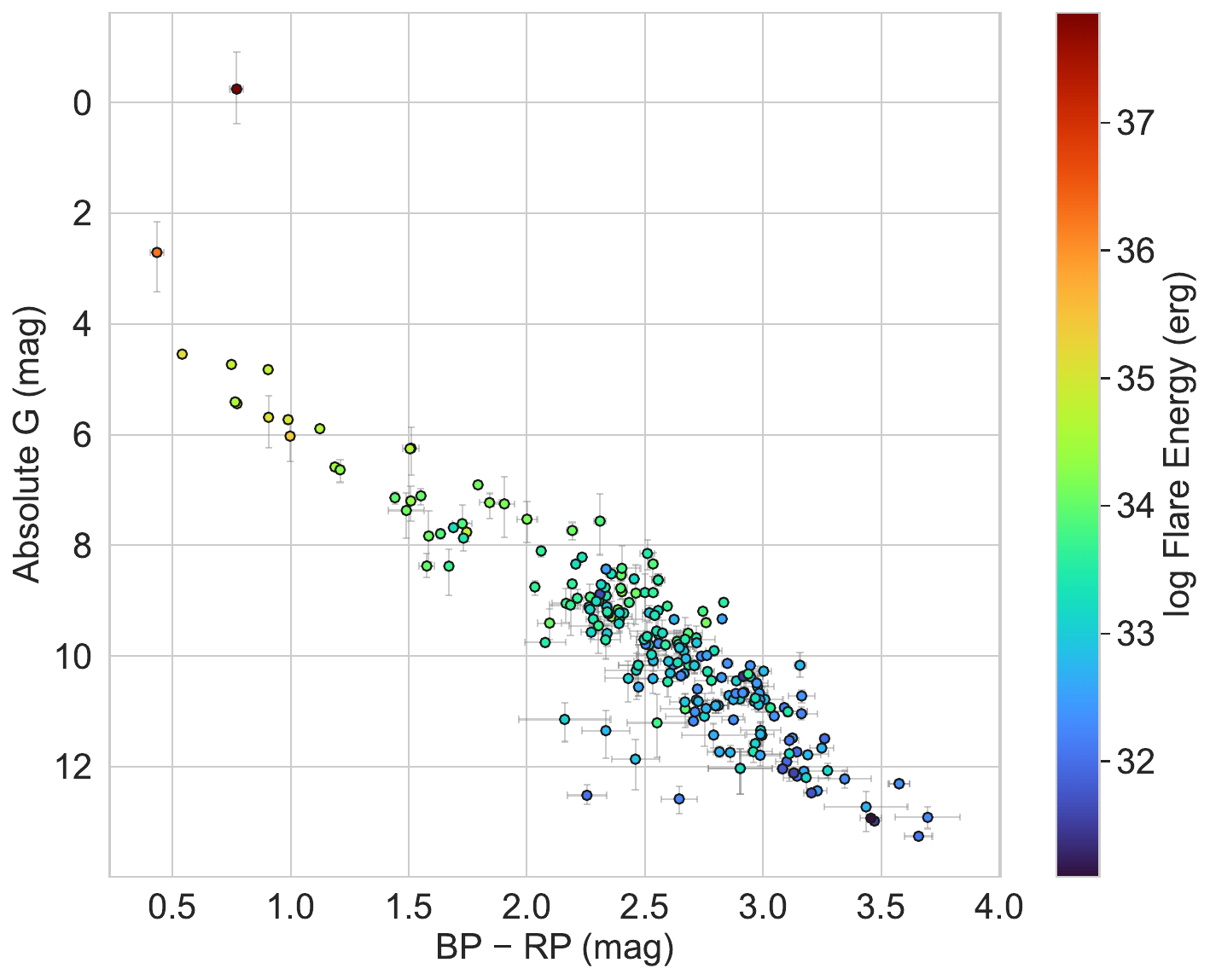}
    \caption{Color Magnitude diagram for all the flaring samples. The color indices and the absolute magnitude are dereddened to take into account the effect of extinction.}
    \label{fig:cmd}
\end{figure}

Figure \ref{fig:cmd} shows the spectral distribution of flaring stars. It shows that flare energy tends to decrease with increasing stellar color (i.e., decreasing effective temperature), while it increases with decreasing absolute magnitude or increasing luminosity. The relation between luminosity and flare energy reflects scaling effects, with larger stars having proportionately larger active regions and thus greater stored magnetic energy \citep{balona_flare_2015}. Color indices and absolute magnitude show weak positive correlations with amplitude and weak negative correlations with log flare duration. 

\subsection{Flare Morphology}
\label{sec:morph}

\begin{deluxetable*}{l c c c}[!t]
\tablecaption{Flare Morphology Counts by Luminosity Class \label{tab:morph_lum}}
\tablehead{
\colhead{Luminosity Class} & \colhead{Classical} & \colhead{Peak-bump} & \colhead{Multi peak Complex}
}
\startdata
Hot Main Sequence & 2 & 3 & 4 \\
Cold Main Sequence I & 90 & 28 & 59 \\
Cold Main Sequence II & 60 & 17 & 38 \\
Sub giant/Red giants & 5 & 0 & 3 \\
\enddata
\tablecomments{Numbers presented here describe the distribution of flare morphologies across luminosity classes. The multi peak complex morphology represents flares with more than 2 peaks as fitted by the modeling algorithm presented in Section \ref{sec:model_fl}.}
\end{deluxetable*}

Based on the model fits obtained using the method outlined in Section \ref{sec:model_fl}, we have classified the flares into three broad categories: 
\begin{itemize}
    \item Classical flares: Characterized by a single peak.
    \item Peak-bump flares: Exhibiting two peaks; a primary peak and an associated secondary peak that appears either during the rising phase or in the decaying tail of the flare.
    \item Multi peak Complex flares: Associated with multiple peaks, indicating more intricate energy release processes.
\end{itemize}

\begin{deluxetable}{l c c c}[!t]
\tablecaption{Flare Morphology Counts by Spectral Type \label{tab:morph_spectral}}
\tablehead{
\colhead{Spectral Type} & \colhead{Classical} & \colhead{Peak-bump} & \colhead{Multi peak Complex}
}
\startdata
M & 117 & 30 & 67 \\
K & 26 & 12 & 24 \\
G & 6 & 3 & 6 \\
F & 2 & 1 & 3 \\
A & 1 & 1 & 1 \\
\enddata
\tablecomments{This table presents flare morphology distribution across spectral types. The multi peak complex morphology represents flares with more than 2 peaks as fitted by the modeling algorithm presented in Section \ref{sec:model_fl}.}
\end{deluxetable}

We have identified 159 classical flares, 105 peak-bump flares, and 50 other complex flares. Therefore, our sample has 49.4\% complex flares (including peak-bump and multi peak) and 50.6\% classical flares. These occurrence rates are comparable to other flare morphology studies where \citet{howard_no_2022} found 42.3\% of 20 s TESS sample to be complex flares, with $\sim70\%$ of complex flares having energy greater than $10^{35}$ erg. \citet{bruno_detailed_2024} found that $\gtrsim30\%$ of their high cadence sample comprises of complex flare while \citet{yudovich_analyzing_2025} found that 61\% of their 10142 stars have complex profile. Tables \ref{tab:morph_lum} and \ref{tab:morph_spectral} show the counts of flares across different luminosity classes and spectral types. We find that M dwarfs show 9.5\% occurrence rate for peak-bump flares, which is higher than the values reported by \citet{howard_no_2022}(7\%) and \citet{yudovich_analyzing_2025}(1.0\%). Additionally, K dwarfs show an occurrence rate of 3.7\% for peak-bump flares in our samples. Figure \ref{fig:morph_en} shows the distribution of energy across different morphologies. We find that among the complex flares, $\sim71.2\%$ have energy greater than $10^{35}$ erg consistent with \citet{howard_no_2022}. We find that complex flares tend to have slightly higher typical energies than classical flares, consistent with the interpretation that complex flares involve more intricate magnetic reconnection processes, thereby releasing greater energy. 

\begin{deluxetable}{l c c c}[!t]
\tablecaption{Flare Morphology Counts by Sub Spectral Type for M dwarfs \label{tab:morph_subspectral}}
\tablehead{
\colhead{Spectral Type} & \colhead{Classical} & \colhead{Peak-bump} & \colhead{Multi peak Complex}
}
\startdata
M0 & 9 & 1 & 5 \\
M1 & 2 & 2 & 6 \\
M2 & 38 & 10 & 14 \\
M3 & 25 & 3 & 13 \\
M4 & 37 & 13 & 25 \\
M5 & 4 & 1 & 2 \\
M6 & 2 & 0 & 2 \\
\enddata
\tablecomments{This table presents flare morphology distribution across sub spectral types for M dwarf stars. The multi peak complex morphology represents flares with more than 2 peaks as fitted by the modeling algorithm presented in Section \ref{sec:model_fl}.}
\end{deluxetable}

\begin{figure}[]
        \plotone{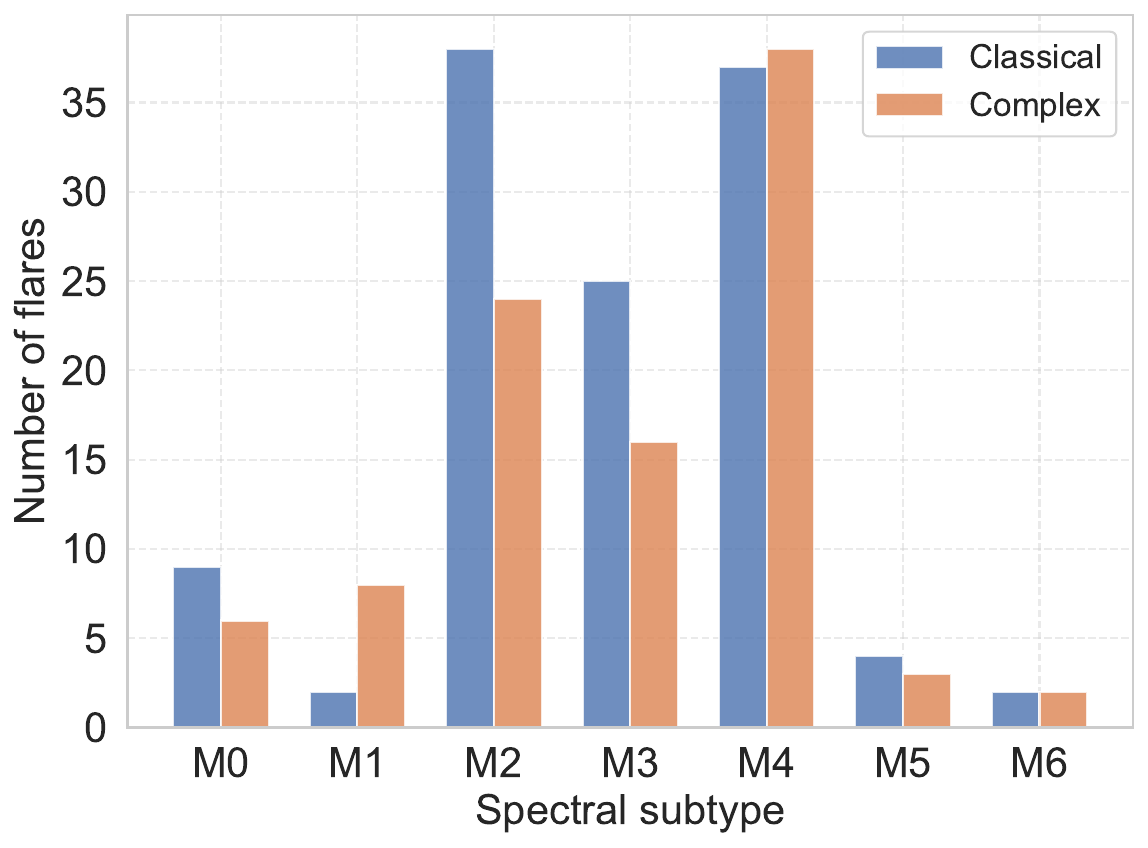}
    \caption{Distribution of flare morphologies across sub spectral types for M dwarfs.}
    \label{fig:morph_2}
\end{figure}

We have further classified flaring M dwarfs into spectral subtype using \citet{pecaut_intrinsic_2013}. Table \ref{tab:morph_subspectral} shows the distribution of flare morphologies across M0-M6. From Figure \ref{fig:morph_2}, we find that M2 and M3 exhibit a clear dominance of the classical flares, while M4 has an equal split between classical and complex morphologies. Earlier (M0-M1) and later (M5-M6) types have too few events for strong statistical statements. The variation in the classical-complex flare ratio across spectral subtypes is dependent on the adopted flare fitting model which is demonstrated in (Appendix \ref{sec:flare1}).

\section{\label{sec:summary}Summary}

In this paper, we have developed an automated algorithm for detecting flares in short-baseline ($\sim$6.2 hours) light curves from the Zwicky Transient Facility (ZTF) Extended Deep-Drilling Survey. This resulted in a catalog of 310 flaring stars across evolutionary stages and spectral types. We have estimated flare energies for a small subset of the flaring stars due to limited distance information. The flare energies span from $10^{31} $ to $ 10^{37}$ ergs, consistent with previous studies. We find a negative correlation between flare energy and absolute magnitude $M_G$ in agreement with the luminosity effect discussed by \citet{balona_flare_2015}. 

We also modeled the flare light curves using the empirical template proposed by \citet{mendoza_llamaradas_2022}, which provides a flexible description of flare morphology through parameters for FWHM, peak time, and amplitude. This allowed us to perform morphological studies of flare energy, revealing that both classical and complex flares exhibit similar distributions. 

Additionally, we showed a power-law relationship between flare energy and duration, with an index of 0.19 ± 0.01. This finding is the lowest estimate in the literature \citep{yang_properties_2023}. A similar positive correlation is found between the decaying time scale and the flare energy, indicating that the cooling mechanism dominates the flare energetics. 

In future work, the methodology developed here can be extended to a broader range of ZTF fields as the survey continues, allowing us to significantly increase the sample size and probe rarer, more extreme flare events. Despite the short duration of Extended Deep Drilling survey, our detection of more than 300 flares in a single night demonstrate the high yield of short cadence surveys. This also highlights the potential of dedicated high cadence campaigns with the Vera C. Rubin Observatory's Legacy Survey of Space and Time \citep[LSST;][]{lsst_2019} to detect and characterize much larger sample of stellar flares. The current method however suffers from a large number of false positives requiring manual vetting; further improvements in automated detection algorithms could enhance the robustness and efficiency of flare identification in even larger datasets.

\begin{acknowledgments}
We acknowledge support from the Department of Atomic Energy, Government of India, under Project
Identification No. RTI 4002.

This work has made use of data from the European Space Agency (ESA) mission Gaia (\url{https://www.cosmos.esa.int/gaia}), processed by the Gaia Data Processing and Analysis Consortium (DPAC, \url{https://www.cosmos.esa.int/web/gaia/dpac/consortium}). Funding for the DPAC has been provided by national institutions, in particular the institutions participating in the Gaia Multilateral Agreement.

This work also used data from ZTF. Based on observations obtained with the Samuel Oschin 48-inch Telescope at the Palomar Observatory as part of the Zwicky Transient Facility project. ZTF is supported by the National Science Foundation under Grant No. AST-1440341 and a collaboration including Caltech, IPAC, the Weizmann Institute for Science, the Oskar Klein Center at Stockholm University, the University of Maryland, the University of Washington, Deutsches Elektronen-Synchrotron and Humboldt University, Los Alamos National Laboratories, the TANGO Consortium of Taiwan, the University of Wisconsin at Milwaukee, and Lawrence Berkeley National Laboratories. Operations are conducted by COO, IPAC, and UW.

We thank the anonymous reviewer for their constructive comments and suggestions, which helped improve the clarity of the manuscript. We also acknowledge the use of ChatGPT (OpenAI) for assistance with language editing.

\end{acknowledgments}

\appendix
\section{\label{sec:app}Impact of Flare Model Flexibility on Morphology Classification}
\label{sec:flare1}

\begin{figure*}
    \gridline{\fig{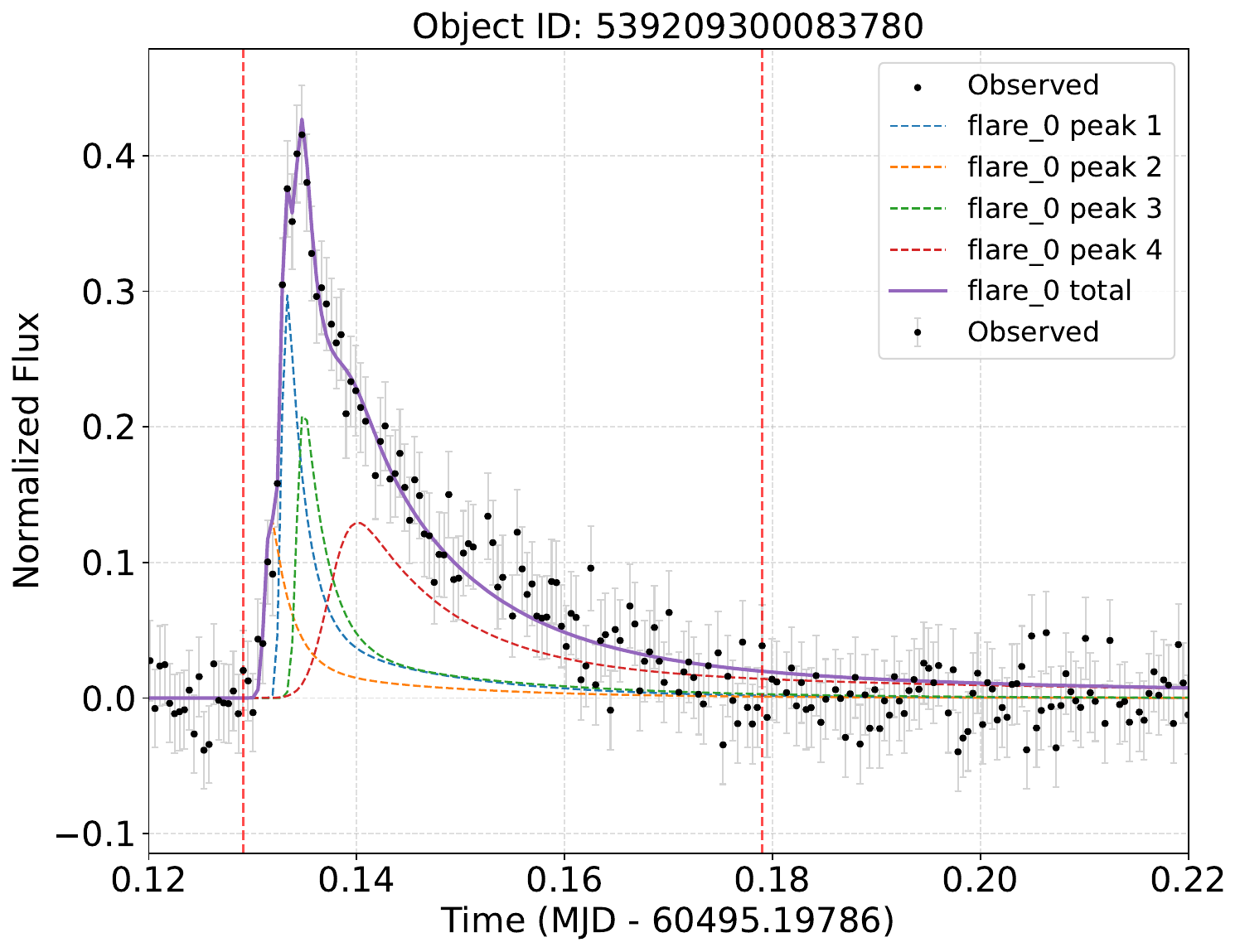}{0.48\textwidth}{(a)}
          \fig{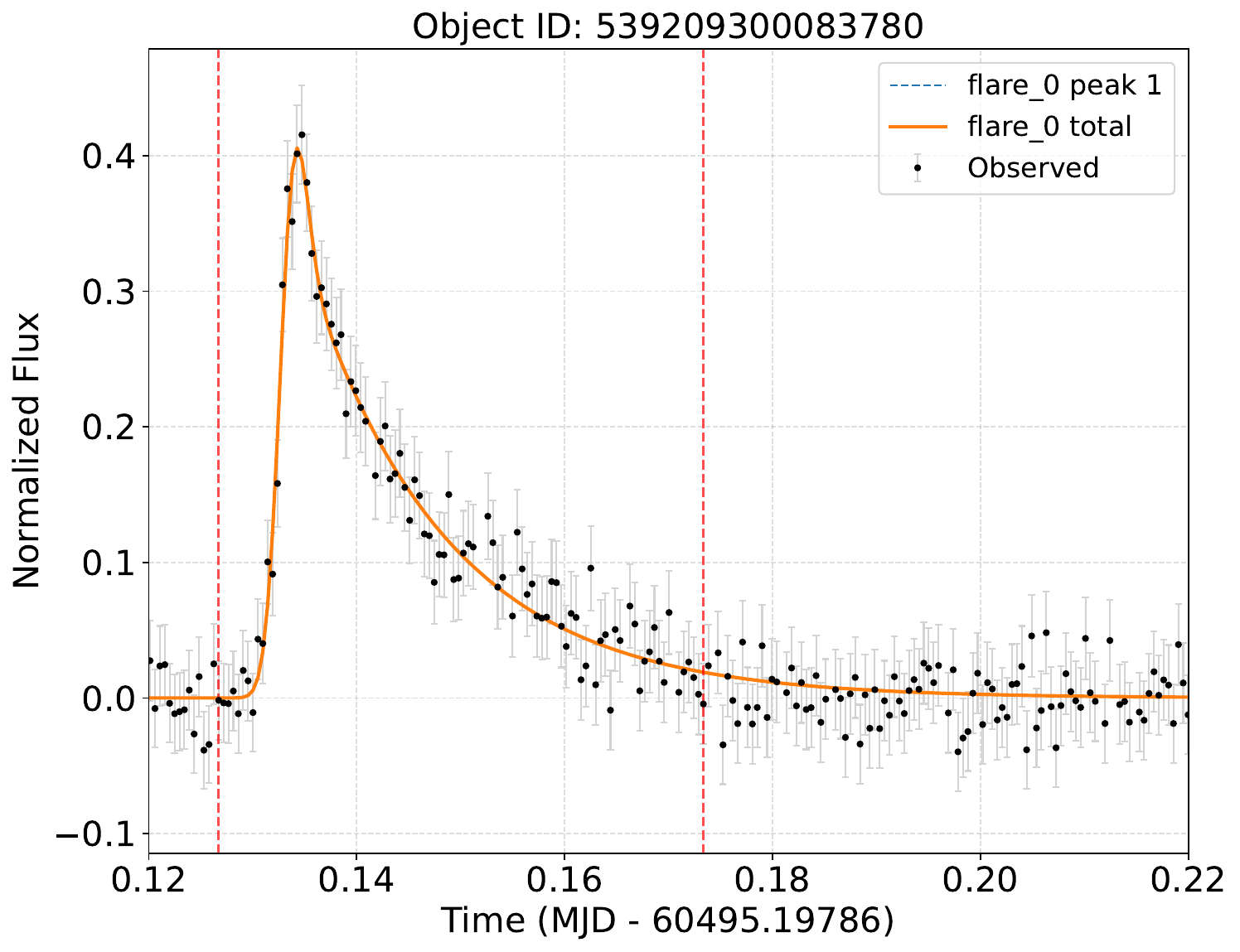}{0.48\textwidth}{(b)}}
    \caption{Left: Flare modeled using the model parameters described in section \ref{sec:model_fl}. The flare is categorized as multi peak complex flare. Right: The same flare modeled by allowing all the parameters to freely vary resulted in the flare to be categorized as a classical flare. The start of the flare from the model is defined as the time when the flux first becomes non-zero whereas the end is defined as the point when flux becomes one-eighth of its peak value.}
    \label{fig:diff_morph}
\end{figure*}

\begin{figure}
        \plotone{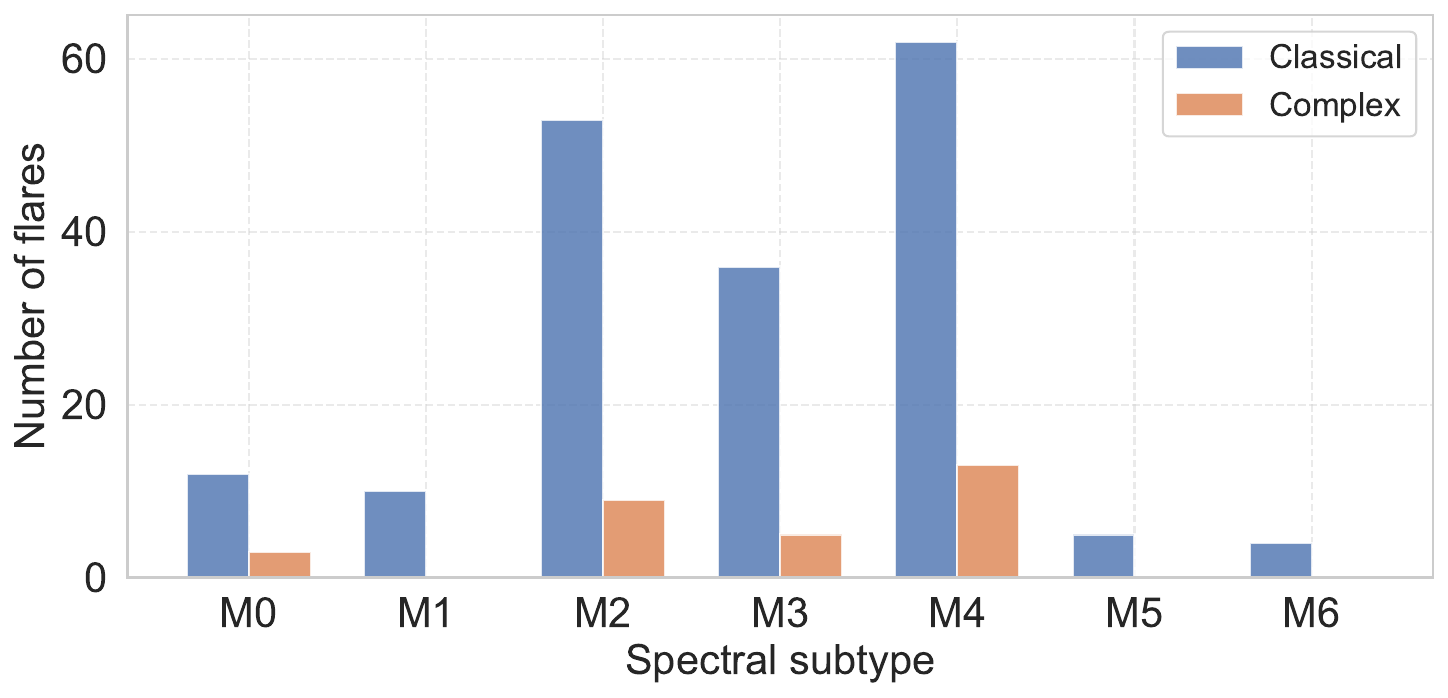}
    \caption{Updated distribution of flare morphologies across sub spectral types for M dwarfs using the new flexible flare model.}
    \label{fig:morph_free}
\end{figure}

In the primary analysis, we adopted the standard assumption common in the literature that stellar flares have a universal analytical shape across luminosity groups and spectral classes. This effectively fixes the shape parameters of the \citet{mendoza_llamaradas_2022} model. 

To assess how strongly this assumption influences the inferred flare properties, we performed an alternative set of fits in which all model parameters: A, B, C, $D_1$, $D_2$, $F_1$, and FWHM, were allowed to vary freely. This breaks the universal-shape assumption, allowing the model to capture a broader range of flare profiles.

We find that freeing the parameters has a negligible effect on the fitted amplitude and flare energy. It does modify the rising and decaying timescales, and consequently the inferred flare FWHM and duration. However, these changes do not alter the overall distributions of flare properties. In addition, we also investigated whether the fitted parameters of the flexible model show any dependencies with stellar parameters such as $BP-RP$ and $M_G$, but we did not find any strong correlation between them.

The most noticeable effect is a shift in the morphology classification, with many flares being labeled as "classical" rather than "complex". One such example is shown in Figure \ref{fig:diff_morph}. This is expected since additional freedom in the flexible model can explain the scattered data with a single broader or asymmetric flare, whereas more components are needed to explain the small scale structures in the restricted model. Although the high cadence data can reveal the small scale structures in the flare profile \citep{howard_no_2022}, ZTF observes extremely faint stars (Figure \ref{fig:hist1}(d)) with substantial photometric uncertainties. Consequently, the restricted model can infer these substructures, but the significance of such structures becomes difficult to establish. This effect is more defined when the peak maxima are comparable within the photometric uncertainties. In such cases, when complex flares are reclassified as simple flares, the flexible model is a better description of the flare with $\Delta \rm AIC = \rm AIC_{\rm restricted}-\rm AIC_{\rm flexible}$ exceeding 15. However, the reclassification of multi peak flare to peak-bump flares results in a small positive $\Delta\rm AIC$, indicating a weaker preference for the flexible model. We report updated counts in Tables \ref{tab:morph_free_lum} and \ref{tab:morph_free_spectral}, and Figure \ref{fig:morph_free} shows the shift toward classical morphologies across spectral subtype.

\begin{deluxetable}{l c c c}[t]
\tablecaption{Updated Flare Morphology Counts by Luminosity Class using the new flexible flare model\label{tab:morph_free_lum}}
\tablehead{
\colhead{Luminosity Class} & \colhead{Classical} & \colhead{Peak-bump} & \colhead{Multi peak Complex}
}
\startdata
Hot Main Sequence & 6 & 3 & 0 \\
Cold Main Sequence I & 150 & 25 & 2 \\
Cold Main Sequence II & 96 & 16 & 1 \\
Sub giant/Red giants & 8 & 0 & 0 \\
\enddata
\end{deluxetable}

\begin{deluxetable}{l c c c}[!h]
\tablecaption{Updated Flare Morphology Counts by Spectral Type using the new flexible flare model \label{tab:morph_free_spectral}}
\tablehead{
\colhead{Spectral Type} & \colhead{Classical} & \colhead{Peak-bump} & \colhead{Multi peak Complex}
}
\startdata
M & 182 & 27 & 3 \\
K & 53 & 9 & 0 \\
G & 10 & 5 & 0 \\
F & 6 & 0 & 0 \\
A & 1 & 2 & 0 \\
\enddata
\end{deluxetable}

We therefore conclude that the universal flare shape model assumption does not affect the derived flare properties. Its impact is limited to the morphology labeling, where a single peak flexible model is able to explain the structure previously interpreted as complexity. The primary results are therefore robust to the assumed analytical form of the flare model.

\bibliography{references}{}
\bibliographystyle{aasjournalv7}

\end{document}